\documentclass[float,epsfig,usenatbib]{mn2e}

\usepackage{graphicx}
\usepackage{appendix}
\usepackage{psfrag}
\usepackage{amsmath,amssymb}
\usepackage{threeparttable}
\usepackage{float}
\usepackage{subfigure}
\usepackage{afterpage}
\usepackage{longtable,lscape}
\usepackage{rotating}

\title[The GALEX Arecibo SDSS Survey. VIII.]{The GALEX Arecibo SDSS
  Survey. VIII. Final Data Release -- The Effect of Group Environment
  on the Gas Content of Massive Galaxies}

\author[B. Catinella et al.]
{Barbara Catinella$^{1,2}$\thanks{bcatinella@swin.edu.au},
  David Schiminovich$^{3}$, Luca Cortese$^{2,4}$, Silvia Fabello$^{5}$, 
  \newauthor Cameron B. Hummels$^{6}$, Sean M. Moran$^{7}$, Jenna J. Lemonias$^{3}$,
  \newauthor Andrew P. Cooper$^{8}$, Ronin Wu$^{9}$, Timothy M. Heckman$^{10}$, and Jing Wang$^{1}$ \\
$^{1}$Max-Planck Institut f\"{u}r Astrophysik, D-85741 Garching, Germany\\
$^{2}$Centre for Astrophysics and Supercomputing, Swinburne University of Technology, Hawthorn, VIC 3122, Australia \\
$^{3}$Department of Astronomy, Columbia University, New York, NY 10027, USA\\
$^{4}$European Southern Observatory, D-85748 Garching, Germany\\
$^{5}$Autoliv Electronics Germany, Theodor-Heuss-Str. 2, 85221 Dachau, Germany\\
$^{6}$Department of Astronomy and Steward Observatory, University of Arizona, Tucson, AZ 85721, USA\\
$^{7}$Smithsonian Astrophysical Observatory, 60 Garden Street, Cambridge, MA 02138, USA\\
$^{8}$National Astronomical Observatories, Chinese Academy of Sciences, 20A Datun Rd., Chaoyang, Beijing 100012, P.R. China\\
$^{9}$Commissariat \`a l'Energie Atomique (CEA), 91191 Gif-sur-Yvette, France\\
$^{10}$Department of Physics and Astronomy, Johns Hopkins University, Baltimore, MD 21218, USA
}

\date{}

\begin{document}

\def\deg{$^{\circ}$}
\newcommand{\eg}{{\it e.g.}}
\newcommand{\ie}{{\it i.e.}}
\newcommand{\minusone}{$^{-1}$}
\newcommand{\kms}{km~s$^{-1}$}
\newcommand{\kmsm}{km~s$^{-1}$~Mpc$^{-1}$}
\newcommand{\Ha}{$\rm H\alpha$}
\newcommand{\Hb}{$\rm H\beta$}
\newcommand{\hi}{{H{\sc i}}}
\newcommand{\hii}{{H{\sc ii}}}
\newcommand{\nii}{\ion{N}{2}}
\newcommand{\rband}{{\em r}-band}
\newcommand{\iband}{{\em I}-band}
\newcommand{\zband}{{\em z}-band}
\newcommand{\rd}{$r_{\rm d}$}
\newcommand{\whi}{$W_{50}$}
\newcommand{\ds}{$\Delta s$}
\newcommand{\x}{$\times$}
\newcommand{\about}{$\sim$}
\newcommand{\Msun}{M$_\odot$}
\newcommand{\Lsun}{L$_\odot$}
\newcommand{\Mhi}{$M_{\rm HI}$}
\newcommand{\Mst}{$M_\star$}
\newcommand{\must}{$\mu_\star$}
\newcommand{\nuvr}{NUV$-r$}
\newcommand{\Rinz}{$R_{50,z}$}
\newcommand{\Ropt}{$R_{\rm opt}$}
\newcommand{\sov}{$S_{0.5}$}
\newcommand{\vrot}{$V_{\rm rot}$}
\newcommand{\vs}{$V_{\rm rot}/\sigma$}
\newcommand{\cindx}{$R_{90}/R_{50}$}
\newcommand{\rhalf}{$R_{50}$}
\newcommand{\tmax}{$T_{\rm max}$}
\newcommand{\ngal}{$N_{\rm gal}$}
\newcommand{\ntot}{$N_{\rm tot}$}
\newcommand{\Mh}{$M_{\rm h}$}
\newcommand{\detfr}{$N_{\rm det}$/$N_{\rm tot}$}
\newcommand{\gi}{$g-i$}

\maketitle

\label{firstpage}

\begin{abstract}

We present the final data release from the GALEX Arecibo SDSS
Survey (GASS), a large Arecibo program that measured the
\hi\ properties for an unbiased sample of \about 800 galaxies with
stellar masses greater than $10^{10}$ \Msun\ and redshifts
$0.025<z<0.05$. This release includes new Arecibo observations for 250
galaxies. We use the full GASS sample to investigate environmental
effects on the cold gas content of massive galaxies at fixed stellar
mass. The environment is characterized in terms of dark matter halo
mass, obtained by cross-matching our sample with the SDSS group
catalog of Yang et al.
Our analysis provides, for the first time, clear statistical evidence
that massive galaxies located in halos with masses of
$10^{13}-10^{14}$ \Msun\ have at least 0.4 dex less \hi\ than 
objects in lower density environments.
The process responsible for the suppression of gas in group galaxies most
likely drives the observed quenching of the star formation in these
systems. Our findings strongly support the importance of the group
environment for galaxy evolution, and have profound implications for
semi-analytic models of galaxy formation, which currently do not allow
for stripping of the cold interstellar medium in galaxy groups.

\end{abstract}


\section{Introduction}\label{s_intro}

As the source of the material that will eventually form stars,
atomic hydrogen (\hi) is clearly a key ingredient to understand how
galaxies form and evolve. For instance, physical processes that
transform galaxies from blue, star-forming to ``red and dead'' objects
must deplete their gas reservoirs first, so that their star formation is quenched as
a result. Systematic studies of the cold gas content of galaxies as a
function of their star formation, mass and structural properties, and across all
environmental densities \citep[\eg][]{gass1,huang12}, are necessary to explain the variety of
systems observed today in the local Universe, and to provide important
constraints to theoretical models and simulations of galaxy formation
\citep[\eg][]{fu10,lagos11,dave11b,coldgass3}.

Environmental mechanisms are known to be effective in
removing gas from galaxies in high-density regions, and indeed \hi\ is one of
the most sensitive tracers of environmental effects. This is because
\hi\ gas typically extends further away from the center of galaxies
compared to other baryonic components, thus it is more
easily affected by environment.
A classic example of the value of \hi\ observations in this context is
represented by spatially-resolved radio observations of the M81 group, which 
have revealed a spectacular, complex network of gas filaments
connecting three galaxies that appear completely undisturbed in optical images
\citep{yun94}.

Despite its importance as environmental probe, we are far from having
a comprehensive picture of how the \hi\ content of galaxies varies as
a function of the local density. This is in stark contrast with
optical studies, where the availability of large photometric and
spectroscopic databases such as those assembled by the Sloan Digital Sky Survey
\citep[SDSS;][]{sdss} and the Two-degree Field Galaxy Redshift 
Survey \citep[2dFGRS;][]{2dF} has allowed
us to quantify how the star formation properties of galaxies vary
across all environments, from voids to clusters, and for different cosmic
epochs \citep[\eg][]{balogh04,gk04,cooper06}.
The evidence based on such datasets suggests that 
the transformation from star-forming to quiescent
galaxies is a smooth function of density, and happens in great part
outside clusters (\eg\ \citealt{dressler80,lewis02,gomez03,blanton09}).
Surprisingly enough, we have not
pinned down the mechanisms that drive this decrease in star formation
rate, and whether this is accompanied/triggered by gas removal. This
is due to a lack of \hi\ observations covering a large enough range of
environments to sufficient depth.

Environmental \hi\ studies to date concentrated on the difference
between cluster and field populations, and demonstrated that galaxies
in high-density regions are \hi\ deficient compared to isolated
objects with similar size and stellar morphology \citep{gh85,solanes01}. 
Resolved \hi\ maps of galaxies in the Virgo and
Coma clusters clearly show that \hi\ is removed from the star-forming disk
\citep{gavazzi89,cayatte90,bravoalfaro00,kenney04,viva}, mainly due to ram pressure
stripping by the dense intracluster medium through which galaxies move
\citep{gunn72,vollmer09,ale_review06}. What happens to the gas content in
the lower density group environment, where ram pressure is thought to
be inefficient, is still unclear. Several studies have mapped the
\hi\ content of galaxies in groups, and found examples of
\hi-deficient galaxies \citep[\eg][]{huchtmeier97,verdes-montenegro01,kilborn09}.
Tidal interactions in groups might funnel gas in the central regions
of galaxies and increase their star formation \citep{iono04,kewley06}, eventually reducing
their \hi\ content, but the net effect on statistical basis is unknown.

Because of limitations in the current \hi\ samples, which target a
limited range of environmental densities, with largely different selection
criteria, \hi\ sensitivities and multi-wavelength coverage, we still do not
know at which density scale the environment starts affecting the gas
content of galaxies. 
In order to quantify the effect of environment on the \hi\ reservoir of
galaxies, we need wide-area surveys over large enough volumes to
sample a variety of environments, and deep enough to probe the
\hi-poor regime. Accompanying multi-wavelength information is
essential not only to determine the environmental density, but also to
provide measurements of the structural and star formation properties
of the galaxies, that are necessary to connect the fate of the gas to
that of the stars. In particular, because star formation and galaxy
properties are known to scale primarily with mass
\citep[\eg][]{gk03,shen03,baldry04}, environmental comparisons must
be done at fixed stellar mass.

\hi-blind surveys such as the ongoing Arecibo Legacy Fast ALFA
\citep[ALFALFA;][]{alfalfa} survey map large volumes, but are not
sensitive enough to detect \hi-poor systems beyond the very local
Universe \citep{gavazzi13}. However, the availability of high-quality
\hi\ spectra for galaxies that are individually not detected can offer important
constraints on the average gas content of galaxies, when these are
binned according to a given property and co-added or ``stacked''
\citep[\eg][]{fabello1,fabello2}.
Indeed, statistical analyses based on stacking of
optically-selected galaxies in the ALFALFA data cubes have already
provided interesting insights into the average \hi\ content of nearby
massive galaxies in groups. \citet{fabello3} found that the average
\hi\ gas mass fraction declines with environmental density, and that
such decline is stronger
than what is observed for the mean global and central specific star
formation rates. By comparing the observed trends with the results of
semi-analytic models, they concluded that ram pressure stripping is
likely to become effective in groups.

In this work, we use deep \hi\ observations of optically-selected
galaxies from the recently completed GALEX Arecibo SDSS Survey 
\citep[GASS;][hereafter DR1]{gass1}
to investigate the effects of the environment on a galaxy-by-galaxy
basis. GASS includes \hi\ measurements for \about 800 galaxies with
stellar masses greater than  $10^{10}$ \Msun\ and redshifts $0.025 < z < 0.05$.
For these galaxies, we have homogeneous measurements of structural
parameters from SDSS and ultraviolet (UV) photometry from GALEX
\citep{galex} imaging. In addition to its clean selection criteria,
GASS is unique for being gas fraction limited: we designed the survey
to reach small limits of gas content at fixed stellar mass 
(\Mhi/\Mst \about 2-5\%), therefore probing
the \hi-rich to \hi-poor regime. Because there is no morphological or
environmental selection, and our redshift cut spans a large volume
(approximately corresponding to distances between 100 and 200 Mpc),
GASS probes a variety of local densities to significant depth, and
thus is ideally suited to investigate environmental effects on the gas
content of massive galaxies.

This paper is organized as follows. We summarize our
survey design and Arecibo observations in Section 2, and 
introduce our third and final data release,
which includes new Arecibo observations for 250 galaxies,
in Section 3 (the catalogs are in Appendix~A).
Sections 4 and 5 illustrate the \hi\ properties of the full GASS
sample and revisit the gas fraction scaling relations introduced in
our earlier work. Section 6 briefly describes the group catalog (based
on SDSS) used to characterize the environment of GASS galaxies, and
presents our results on the environmental analysis. Discussion and
conclusions follow in Section 7. 
All the distance-dependent quantities in this work are computed
assuming $\Omega=0.3$, $\Lambda=0.7$ and $H_0 = 70$ \kmsm. 
AB magnitudes are used throughout the paper.

\section{Sample selection, Arecibo observations and data reduction}\label{s_sample}

Survey design, sample selection, Arecibo observations and data
reduction are described in detail in our first two data release papers
(DR1 and \citealt{gass6}, hereafter DR2),
thus we only provide a summary here, including relevant updates.

GASS was designed to measure the global \hi\ properties of \about 1000 galaxies,
selected uniquely by their stellar mass ($10 < {\rm Log} (M_\star/M_\odot) < 11.5$)
and redshift ($0.025 < z < 0.05$). The galaxies are located
within the intersection of the footprints of the SDSS primary
spectroscopic survey, the GALEX Medium
Imaging Survey and ALFALFA. We defined a GASS {\it parent sample},
based on SDSS DR6 \citep{sdss6} and the final ALFALFA
footprint, which includes 12006 galaxies that meet our survey
criteria. The targets for 21cm observations were chosen 
by randomly selecting a subset of the parent sample which balanced the
distribution across stellar mass and which maximized existing GALEX
exposure time. 

We observed the galaxies with the Arecibo radio telescope until
we detected them or until we reached a limit of a few percent in
gas mass fraction (defined as \Mhi/\Mst\ in this work). Practically, we 
set a limit of $M_{\rm HI}/M_\star > 0.015$ for galaxies with 
${\rm Log} (M_\star/M_\odot) >10.5$, and a constant gas mass limit 
${\rm Log} (M_{\rm HI}/M_\odot) =8.7$ for galaxies with smaller stellar masses. This
corresponds to a gas fraction limit $0.015-0.05$ for the whole sample.
Given the \hi\ mass limit assigned to each galaxy (set by its gas
fraction limit and stellar mass), we computed the observing time,
\tmax, required to reach that value with our observing mode and
instrumental setup. We excluded from our sample any
galaxies requiring more than 3 hours of total integration
time\footnote{
There are a few exceptions (5\% of the sample), represented by
galaxies added for our initial pilot observations or already observed
by one of our follow-up programs with the Hubble Space Telescope or other facilities
\citep{gass5,coldgass1}.
}
(this
effectively behaves like a redshift cut at the lowest stellar masses).
Galaxies with good \hi\ detections already available from ALFALFA and/or the Cornell
\hi\ digital archive \citep[][hereafter S05]{springob05} were not
re-observed. These \hi-rich galaxies are added back to the GASS
observations to make the {\it representative} sample (see
Section~\ref{s_properties}).

GASS observations started in March 2008 and ended in July 2012.
The total telescope time allocation was 1005 hours, of which 
\about 11\% unusable due to radio frequency interference (RFI) or
other technical problems. This third and final data release includes
the observations carried out after March 1st 2011 (420 hours divided into
117 runs). 

The Arecibo observations were carried out remotely in standard
position-switching mode, using the L-band wide receiver and the
interim correlator as a backend. Two correlator boards with 12.5 MHz
bandwidth, one polarization, and 2048 channels per spectrum (yielding
a velocity resolution of 1.4 \kms\ at 1370 MHz before smoothing) were centered
at or near the frequency corresponding to the SDSS redshift
of the target. We recorded the spectra every second with 9-level sampling.

The data reduction, performed in the IDL environment, includes
Hanning smoothing, bandpass subtraction, RFI excision, and flux
calibration. The spectra obtained from each on/off pair are weighted by 1/$rms^2$,
where $rms$ is the root mean square noise measured in the
signal-free portion of the spectrum, and co-added. The two
orthogonal linear polarizations (kept separated up to this point) are
averaged to produce the final spectrum, which is boxcar smoothed,
baseline subtracted and measured as explained in the DR1 paper. 
The instrumental broadening correction for the velocity widths
is described in the DR2 paper (we revised it after DR1, as discussed
in \citealt{gass4}).

\section{Data release}\label{s_data}

This data release is incremental over DR1 and DR2, and includes new
Arecibo observations of 250 galaxies. The catalogs of optical, UV
and 21 cm parameters for these objects are presented in Appendix~A.

All the optical parameters were obtained from the SDSS
DR7 database server\footnote{
{\em http://cas.sdss.org/dr7/en/tools/search/sql.asp}
}. Stellar masses are from the Max
Planck Institute for Astrophysics (MPA)/Johns Hopkins
University (JHU) value-added catalogs based on SDSS
DR6, and assume a \citet{chabrier03} initial mass function.

The GALEX UV photometry for our sample was reprocessed by
us, as explained in \citet{jing10} and summarized in the DR1 paper.
Briefly, we produced \nuvr\ images by registering GALEX and SDSS
frames, and convolving the latter to the UV point spread
function. The measured \nuvr\ colors are corrected for Galactic
extinction only; we do not apply internal dust attenuation corrections.

The catalogs presented in our three releases are available both
individually and combined on the GASS website\footnote{
{\em http://www.mpa-garching.mpg.de/GASS/data.php}
}, along with all the \hi\ spectra in digital format.

\begin{figure*}
\begin{center}
\includegraphics[width=16cm]{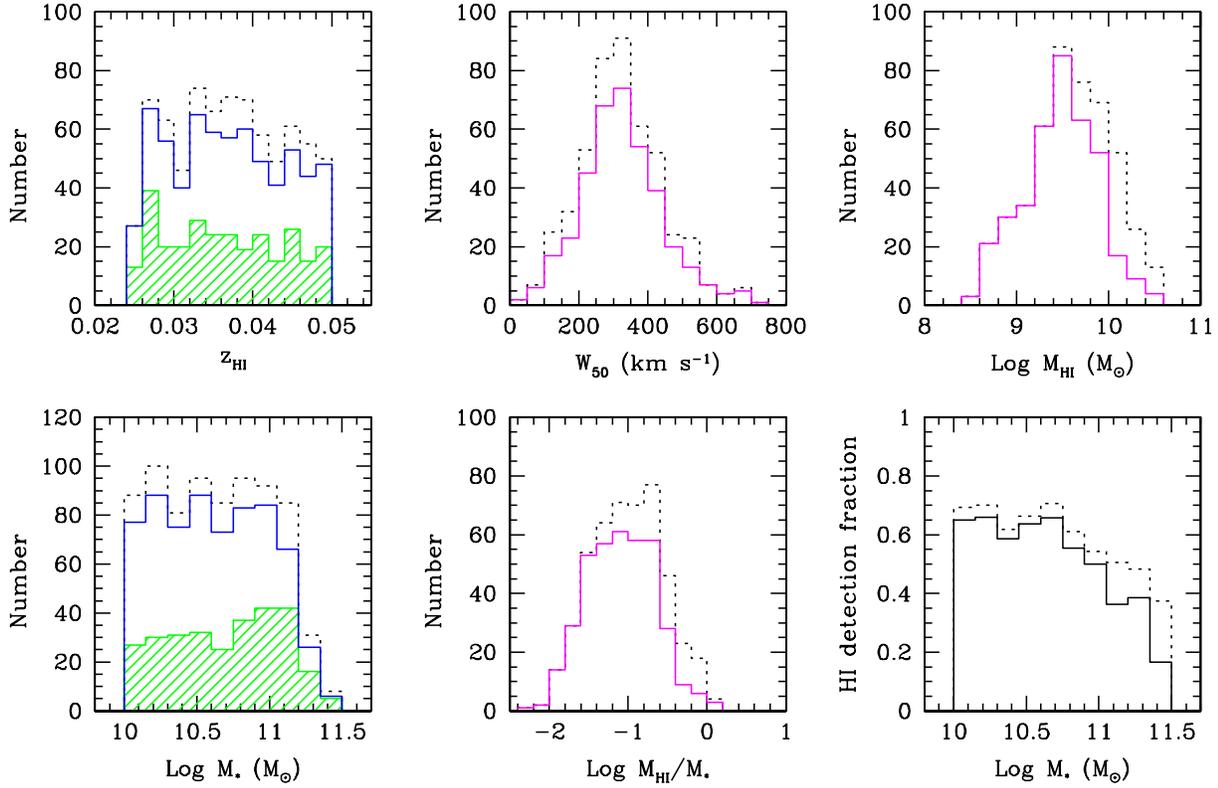}
\caption{GASS sample properties. {\it Top row:} Distributions of \hi\ 
redshifts, velocity widths (not corrected for inclination) and
\hi\ masses for the Arecibo detections (magenta histograms).
The green hatched histogram in the left panel shows the
distribution of SDSS redshifts for the non-detections (the blue
histogram includes both detections and non-detections).
{\it Bottom row:} Distributions of stellar mass (same color scheme as
top left panel), gas fraction, and detection fraction (\ie, the ratio of detections to total)
as a function of stellar mass. The dotted histograms in all
panels correspond to the representative sample, which
includes gas-rich objects from ALFALFA and/or S05 archive (see text).
}
\label{hi}
\end{center}
\end{figure*}

\section{GASS sample properties}\label{s_properties}

The three GASS data releases combined include 666 galaxies,
of which 379 are \hi\ detections and 287 are
non-detections. We refer to this as the GASS {\it observed}
sample. Because we did not reobserve galaxies with good \hi\ detections
already available from either ALFALFA or the S05 archive, this sample
lacks the most gas-rich objects, which need to be added back 
in the correct proportions. By following the procedure
described in the DR1 paper, we obtained a sample that
includes 760 galaxies (of which 473 are detections) and that is
representative in terms of \hi\ properties. We 
refer to this as the GASS {\it representative} sample. Notice that,
because of the improved statistics compared to DR1, here
we use only one such representative sample (as opposed to a
suite of 100 realizations with different sets of randomly-selected
gas-rich galaxies added to the GASS observations).

The \hi\ properties of the detected galaxies are illustrated in
Figure~\ref{hi} for both observed ({\bf solid} histograms) and
representative (dotted) samples. The blue histogram in the top
left panel shows the redshift distribution for the full GASS observed
sample, using the SDSS redshifts for the non-detections (hatched green
histogram). We note that \hi\ detections and non-detections
present a similar redshift distribution.
As for our previous data releases, the distribution of corrected
velocity widths (which have not been deprojected to edge-on view)
peaks near 300 \kms, which is the value that we assume to compute
upper limits for the \hi\ masses of the non-detections, and to
estimate \tmax\ in Table~\ref{t_sdss}. 
The bottom left panel shows the stellar mass distribution for the
observed and representative samples. The corresponding distribution
for the non-detections is shown as a hatched green histogram (as for the
redshift distribution, the detections are plotted on top of the
non-detections). The stellar mass histogram is almost flat by survey
design, as we wish to obtain similar statistics in each bin in order
to perform comparisons at fixed stellar mass. As already noted
in the DR1 and DR2 papers, non-detections span the entire range of stellar masses,
but they are concentrated in the red portion of the \nuvr\ space (not shown).
The detection fraction, \ie\ the ratio of detected galaxies to total,
is plotted as a function of stellar mass in the bottom right panel. The detection
fraction is close to 70\% for \Mst $< 10^{10.7}$ \Msun, and drops to 
\about 40\% in the highest stellar mass bin.

\begin{figure*}
\includegraphics[width=15.5cm]{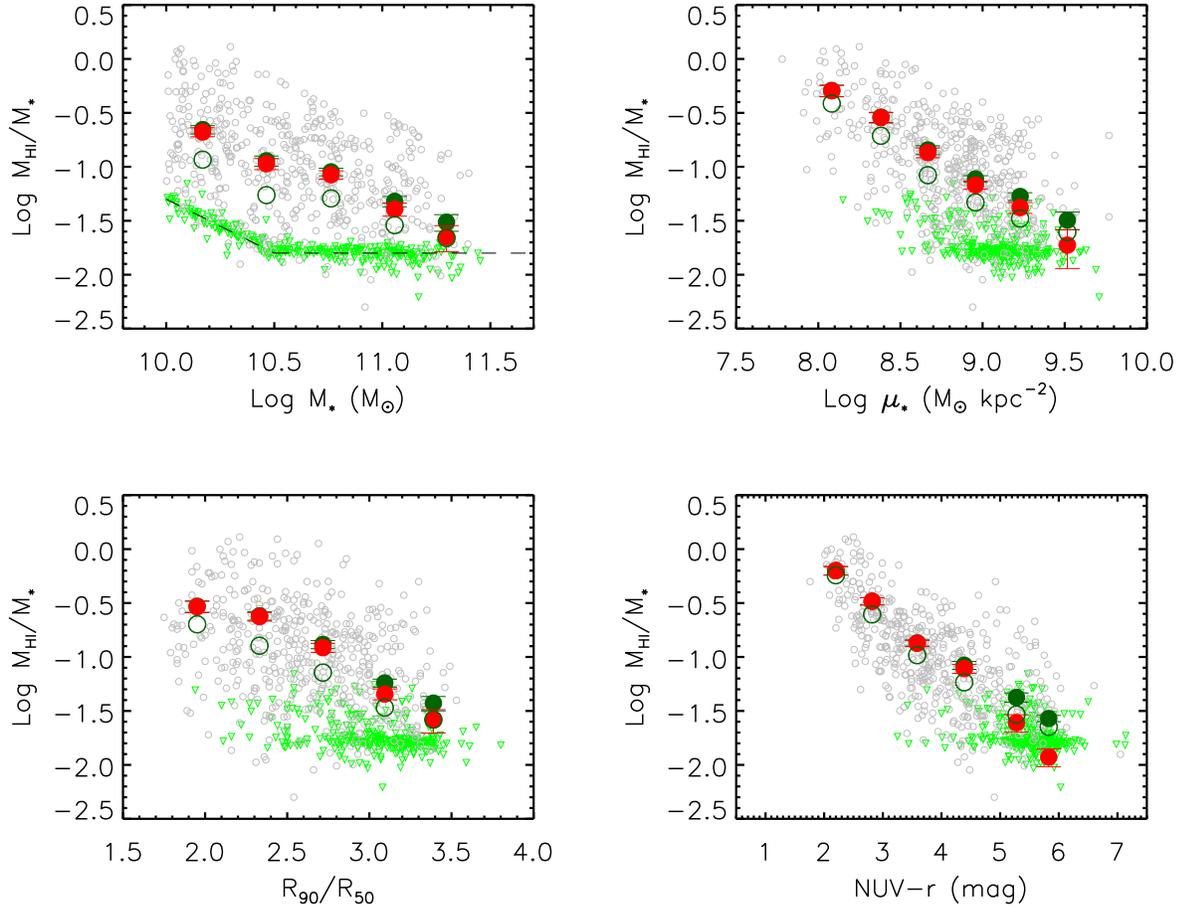}
\caption{Average trends of \hi\ mass fraction as a function of stellar
mass, stellar mass surface density, concentration index and observed
\nuvr\ color for the representative sample. 
In each panel, large filled circles indicate weighted average gas fractions 
(see text). These were computed including the
non-detections, whose \hi\ mass was set to either its upper limit
(green) or to zero (red). Large empty circles indicate weighted
averages of the logarithms of the gas fractions. Only bins including
at least 10 galaxies are shown. These results are listed in Table~\ref{t_avgs}. 
Small gray circles and green upside-down triangles indicate individual
\hi\ detections and non-detections (plotted at their upper limits), respectively.
The dashed line in the top left panel shows the \hi\ gas fraction limit of GASS.}
\label{scalings}
\end{figure*}

\section{Gas fraction scaling relations}\label{s_gfplane}

In this section we present the final version of the scaling relations
introduced in the DR1 paper, now based on the full GASS sample.
Here and in the rest of this work we use the representative
sample for our analysis (unless explicitly noted).

Clockwise from the top left, Figure~\ref{scalings} shows how the gas mass
fraction \Mhi/\Mst\ depends on stellar mass, stellar mass surface
density (defined as $\mu_\star = M_\star/(2 \pi R_{50,z}^2)$, where \Rinz\ is the 
radius containing 50\% of the Petrosian flux in \zband, expressed in kpc units),
observed \nuvr\ color and \cindx\ concentration index (a
proxy for bulge-to-total ratio).
Small gray circles and green upside-down triangles
indicate \hi\ detections and non-detections (plotted at their upper
limits), respectively. 
The average values of the gas fraction are overplotted as filled
circles; these are computed including the
non-detections, whose \hi\ masses were set either to their upper
limits (green) or to zero (red). The
averages are weighted in order to compensate for the flat stellar mass
distribution of the GASS sample, using the volume-limited parent
sample as a reference. Briefly, we binned both parent 
and representative samples by stellar mass (with a 0.2 dex
step), and used the ratio between the two histograms as a weight.
Error bars indicate the standard deviation of the weighted averages.
These results are entirely consistent with our previous findings 
(see also \citealt{fabello1} and \citealt{luca11}). In summary:\\ 
-- The gas fraction of massive galaxies anticorrelates with all the
quantities shown in Figure~\ref{scalings}. The tightest correlations
are with observed \nuvr\ color (Pearson correlation coefficient
$r=-0.69$) and stellar mass surface density ($r=-0.56$), and the weakest
ones are with stellar mass ($r=-0.44$) and concentration index
($r=-0.38$).\\
-- The non-detections are almost exclusively found at 
stellar mass surface densities \must $>10^{8.5}$ \Msun~kpc$^{-2}$ and 
\nuvr\ $>4.5$ magnitudes. The average gas fractions are
insensitive to the way we treat the non-detections, except for the
very most massive, dense and red galaxies.

We chose to compute averages of the linear gas fractions and plot
their logarithms because this allows us to bracket the possible 
\hi\ masses of the non-detections (between zero and their upper
limits). However, as noted by \citet{luca11}, the distribution of
\hi\ gas fraction is closer to lognormal than Gaussian, hence
averaging the logarithms seems more appropriate. In this case
we can only set the non-detections to their upper limits, and the
resulting weighted averages of the logarithmic gas fractions
are plotted in Figure~\ref{scalings} as empty green circles.
These are systematically smaller than the averages of the linear gas
fractions (filled green circles), and the difference is larger for the
stellar mass and concentration index relations, which are also the
most scattered.
The values of the weighted average gas fractions shown in this figure
are listed in Table~\ref{t_avgs} for reference.

\begin{table*}
\centering
\caption{Weighted Average Gas Fractions}
\label{t_avgs}
\begin{tabular}{lrcccr}
\hline\hline
$x$  & $\langle x \rangle$ & $\langle M_{\rm HI}/M_\star \rangle^{a}$  & $\langle M_{\rm HI}/M_\star \rangle^{b}$   & $\langle log (M_{\rm HI}/M_\star) \rangle^{c}$ & $N^{d}$ \\
\hline
Log \Mst  &  10.17 &  0.221$\pm$0.020 &  0.210$\pm$0.020 & $-$0.934 &  188 \\
   	  &  10.46 &  0.114$\pm$0.012 &  0.107$\pm$0.013 & $-$1.262 &  176 \\
   	  &  10.76 &  0.090$\pm$0.007 &  0.084$\pm$0.008 & $-$1.293 &  180 \\
   	  &  11.06 &  0.048$\pm$0.005 &  0.041$\pm$0.006 & $-$1.540 &  177 \\
   	  &  11.30 &  0.031$\pm$0.005 &  0.022$\pm$0.006 & $-$1.660 &	39 \\
          &        &  		      &            	 &   	    &	   \\ 
Log \must &   8.08 &  0.509$\pm$0.061 &  0.508$\pm$0.062 & $-$0.414 &	32 \\
   	  &   8.38 &  0.289$\pm$0.031 &  0.286$\pm$0.031 & $-$0.712 &	69 \\
   	  &   8.67 &  0.143$\pm$0.014 &  0.135$\pm$0.014 & $-$1.078 &  145 \\
   	  &   8.96 &  0.077$\pm$0.005 &  0.068$\pm$0.006 & $-$1.330 &  268 \\
   	  &   9.23 &  0.053$\pm$0.004 &  0.042$\pm$0.005 & $-$1.480 &  218 \\
          &   9.52 &  0.032$\pm$0.006 &  0.019$\pm$0.007 & $-$1.605 &	24 \\
          &        &  		      &            	 &   	    &	   \\ 
\cindx	  &   1.95 &  0.295$\pm$0.035 &  0.293$\pm$0.036 & $-$0.697 &	50 \\
   	  &   2.33 &  0.240$\pm$0.021 &  0.237$\pm$0.022 & $-$0.896 &  160 \\
   	  &   2.72 &  0.131$\pm$0.011 &  0.122$\pm$0.012 & $-$1.145 &  217 \\
   	  &   3.09 &  0.057$\pm$0.005 &  0.045$\pm$0.005 & $-$1.469 &  272 \\
   	  &   3.39 &  0.037$\pm$0.006 &  0.026$\pm$0.006 & $-$1.580 &	60 \\
          &        &  		      &            	 &   	    &	   \\ 
\nuvr 	  &   2.20 &  0.632$\pm$0.057 &  0.632$\pm$0.057 & $-$0.242 &	24 \\
   	  &   2.82 &  0.329$\pm$0.025 &  0.329$\pm$0.025 & $-$0.605 &  108 \\
   	  &   3.59 &  0.135$\pm$0.009 &  0.134$\pm$0.009 & $-$0.983 &  139 \\
   	  &   4.39 &  0.084$\pm$0.007 &  0.078$\pm$0.007 & $-$1.235 &  131 \\
   	  &   5.28 &  0.042$\pm$0.004 &  0.025$\pm$0.005 & $-$1.533 &  194 \\
   	  &   5.83 &  0.027$\pm$0.002 &  0.012$\pm$0.002 & $-$1.648 &  145 \\
\hline\hline
\end{tabular}
\begin{flushleft}
Notes. --- $^{a}$Gas fraction weighted average; \hi\ mass of non-detections set to upper limit.
$^{b}$Gas fraction weighted average; \hi\ mass of non-detections set to zero.
$^{c}$Weighted average of logarithm of gas fraction; \hi\ mass of non-detections set to upper limit.
$^{d}$Number of galaxies in the bin.
\end{flushleft}
\end{table*}

\begin{figure*}
\includegraphics[width=16cm]{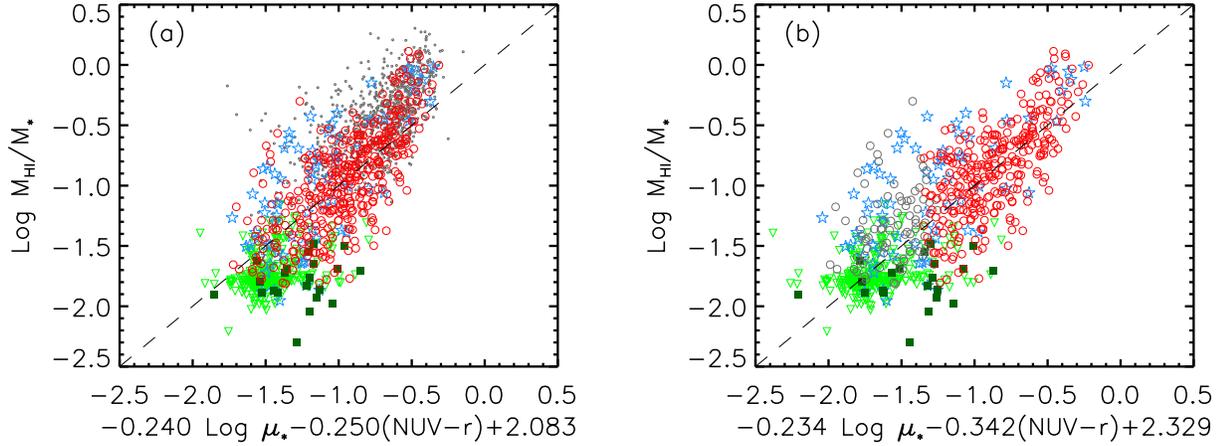}
\caption{Gas fraction plane, a relation between \hi\ mass fraction and a
linear combination of stellar mass surface density and observed
\nuvr\ color. $(a)$ Relation obtained using all the \hi\ detections in the
GASS representative sample (red circles) that are not confused (blue
stars) or below the nominal gas fraction limit of GASS
(dark green squares). Green upside-down triangles are non-detections,
and galaxies meeting GASS selection criteria that have been
cataloged by ALFALFA to date are shown as gray dots. 
$(b)$ Relation obtained using only the subset of detected galaxies with
\nuvr $\leq 4.5$ mag (red circles). Gray circles indicate the
remaining \hi\ detections; green and blue symbols are as in (a).}
\label{plane}
\end{figure*}

In our past work we introduced the {\it gas fraction plane}, a relation
between gas mass fraction and a linear combination of \nuvr\ color
(which is a proxy for star formation rate per unit stellar mass) and
stellar mass surface density, which can be used to define what is
``\hi\ normalcy'' for local massive, star-forming galaxies. 
The plane is obtained by fitting only the \hi\ detections and
minimizing the scatter on the $y$ coordinate (thus, it is
equivalent to a direct fit).
As demonstrated by \citet{luca11}, the distance from the plane along the
$y$ axis strongly correlates with the \hi\ deficiency parameter
\citep{haynes84} and
has a similar scatter (naturally, the sample used to define the plane
should be representative of unperturbed systems). This makes 
the gas fraction plane a very useful tool to investigate environmental
effects and to identify unusually \hi-rich galaxies, especially
when an accurate morphological classification is not available.

We plot the gas fraction plane in 
Figure~\ref{plane}a. We refined our sample by excluding galaxies for
which confusion within the Arecibo beam is certain (because their measured
\hi\ flux belongs entirely or for the most part to a
companion galaxy; these objects are marked as blue stars)
and galaxies with measured gas fractions below our survey limit\footnote{
Figure~\ref{scalings} (top left panel) shows a few \hi\ detections
below the nominal gas fraction limit of GASS (dashed line). As 
explained in the DR1 paper (footnote 6), the main reasons for this
are that (i) the expected gas fraction limit assumes a 5$\sigma$ signal with
velocity width of 300 \kms\ (hence galaxies with smaller widths and/or
face-on might be detected with higher signal to noise), and (ii) we
never integrate less than 4 minutes (but, at large stellar masses, the 
gas fraction limit can be reached in as little as 1 minute).}
(squares). 
For comparison, we also show the full set of ALFALFA galaxies meeting
GASS selection criteria that have been cataloged to date
\citep[][gray dots]{alfalfa40}, and that comprise the most \hi-rich
systems in the GASS volume.
The coefficients of the gas fraction plane are noted on the $x$ axis
of the figure. These have slightly changed with respect to the DR2
version (Log \Mhi/\Mst $=-0.338$ Log \must\ $-0.235$ \nuvr\ $+2.908$), but the two
solutions are entirely consistent: the mean difference between the two
gas fraction predictions is $-0.023$ dex, with a standard deviation of
0.027 dex. The rms scatter of the plane in Log \Mhi/\Mst\ is now 0.292 dex
(it was 0.319 dex for DR2).

As discussed in the DR2 paper, the validity of the gas fraction plane 
breaks down in the region where the contribution of the \hi\
non-detections (which are excluded from the sample used to define it)
becomes significant. Therefore we computed another gas
fraction plane relation using only galaxies with \nuvr $\leq 4.5$ mag,
which is presented in Figure~\ref{plane}b. Over its interval of
validity, this relation has slightly smaller
scatter (0.281 dex) than our original plane in (a). The relation in (b)
should be preferred to predict gas fractions of
massive galaxies on the star-forming sequence. In any other case we 
recommend to use the relation in (a) because it is based on the full
sample of detections, rather than on a subset, and spans the entire
range of \nuvr\ colors and stellar mass surface densities covered by
massive galaxies.

In summary, the average scaling relations have not significantly changed with
respect to our previous data releases, except for the fact that the
errorbars are of course smaller. However, we can now take advantage of
our increased statistics to investigate second order effects, such as
the dependence of the gas content on the environment {\it at fixed
stellar mass}, which would not be feasible without the full survey
sample.

\begin{figure*}
\includegraphics[width=15.5cm]{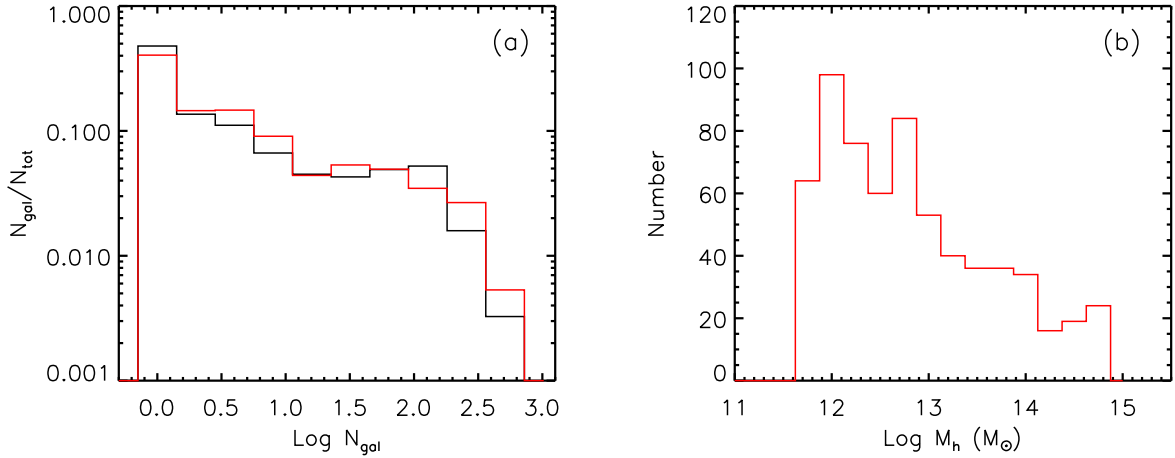}
\caption{$(a)$ Normalized distribution of \ngal, the number of galaxies
in each group, for the GASS parent and representative samples (black
and red respectively; see text). $(b)$ Distribution of halo masses for the
representative sample. This histogram does not include 110 galaxies in
very small groups that do not have halo masses assigned in
the group catalog.}
\label{env_hist}
\end{figure*}

\section{Effect of environment on the gas content of massive galaxies}\label{s_env}

\subsection{Group catalog and halo masses}\label{s_groupcat}

Here we describe briefly the group catalog that we used to
characterize the environment of GASS galaxies.

\citet{yang07} compiled a catalog of galaxy groups based on SDSS DR4,
using what they refer to as a halo-based group finder. Their algorithm is
iterative and includes the following steps: (a) identify
potential group centers using two methods; (b) compute the
characteristic luminosity of each tentative group (\ie\ the combined
luminosity of all group members brighter than a threshold); (c)
estimate the mass, size and velocity dispersion of the dark matter
halo associated with it (initially using a constant mass-to-light
ratio for all groups); (d) reassign galaxies to each tentative group
based on its halo properties; (e) recompute group centers and iterate 
until there is no further change in the group memberships. 

Once the group catalog was finalized, \citet{yang07} assigned halo
masses via abundance matching, assuming the halo mass function of
\citet{warren06}. In practice, they associated the characteristic
luminosity or stellar mass of a group to a halo mass by matching their
rank orders.

They applied the same algorithm to SDSS DR7 \citep{yang12}, and generated two sets of
group catalogs\footnote{
Available at {\it http://gax.shao.ac.cn/data/Group.html}}, one based on Petrosian magnitudes and one based on
model magnitudes. We use the latter for our environmental analysis,
and adopt halo masses \Mh\ obtained by rank ordering the groups by stellar
mass, following \eg\ \citet{woo13}. The catalog also classifies
galaxies as centrals or satellites. 

We note that 10 out of 760
galaxies in our GASS representative sample are not included in the
group catalog, and are thus excluded from our environmental analysis.
Lastly, very small groups are not assigned halo masses
in the group catalog, and this affects 110 of the remaining galaxies.
However, this is not an issue for our analysis, as we will divide
our sample into three intervals of halo mass, and include those 110
galaxies in the lowest \Mh\ bin (Log \Mh/\Msun $<12$).

\subsection{The environment of GASS galaxies}

We begin our analysis by asking what are the typical environments
probed by the GASS galaxies. In order to establish this, we
crossed-matched both our parent and representative samples with the
galaxies in the group catalog described above.
We remind the reader that the parent sample
is the super-set of all the 12,006 galaxies in SDSS DR6 that
meet the GASS selection criteria (stellar mass, redshift cuts and
located within the final ALFALFA footprint), out of which we extracted
those that we observed with Arecibo. As such, the parent sample is
volume-limited and reasonably complete in stellar mass above $10^{10}$
\Msun\ (aside from SDSS fiber collision issues).

We plot the normalized distribution of \ngal, the number galaxies in
each group, in Figure~\ref{env_hist}a, for both parent (black) and
GASS (red) samples. Galaxies with \ngal $=1$ are isolated, and we
generically call ``group'' any structure with two or more members.
According to this definition, about half of the GASS parent sample
galaxies are isolated (48\%; the percentage is 43\% for the
representative sample), and about half are in groups. The richest
structure in our survey volume is represented by the far outskirts of
the Coma cluster (with \ngal $=623$; with a median redshift of 0.0229,
the center of Coma is just below our redshift cutoff).
Compared with the parent sample, the GASS sample probes the
same environments in terms of group richness.
The distribution of halo masses for the GASS sample is shown in panel
(b); the 110 galaxies in small groups mentioned above, which do not
have halo masses assigned in the group catalog, are not plotted.
As a result of our survey strategy (specifically, the fact that
we selected a set of galaxies that balanced the distribution across
stellar mass), this histogram is less peaked at low \Mh\ than the
corresponding one for the parent sample (not shown), but most
importantly the two samples span the same interval of halo mass.

\begin{figure}
\includegraphics[width=8cm]{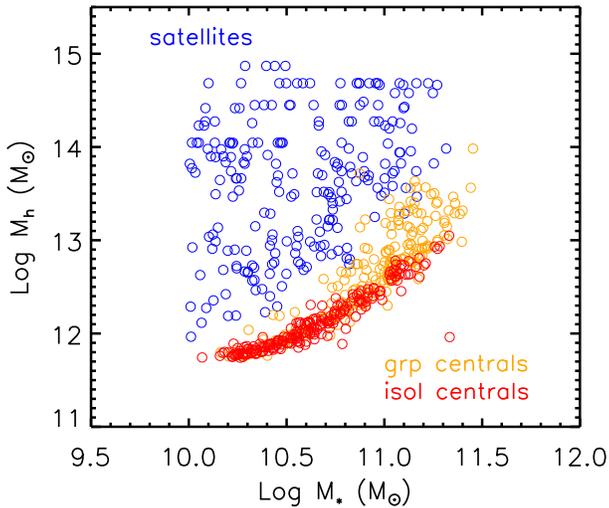}
\caption{Relation between halo and stellar masses for central galaxies
in isolation (red circles) or in groups (orange) and for satellite
galaxies (blue) in the GASS sample.
}
\label{mst_mh}
\end{figure}

Lastly, Figure~\ref{mst_mh} shows the relation between stellar and
halo masses for the galaxies in our sample with assigned halo mass; we
color-coded the points to indicate central galaxies in isolation (red) or
in groups (orange) and satellites (blue). Our sample does not include
central galaxies in the most massive halos, because such systems are
rare.

Having established that the GASS sample is representative of the
parent sample also in terms of environment, it is important to note
that our survey only probes low to intermediate density environments (as we
discuss later, our most massive halo bin is dominated in number by
groups with an average of 20 members). There
are no rich clusters, such as Virgo or Coma, in our survey volume. 
Both the limited dynamic range in galaxy density and our relatively
high gas fraction limit (see below) do not allow us to investigate the
most dramatic cases of \hi\ stripping, which
are well known to occur in the central regions of
clusters and rich groups \citep{cayatte90,bravoalfaro00,ale_review06}.
The reader should bear this in mind when
interpreting our results in the following sections. Instead, GASS is
optimally suited to look for evidence of quenching mechanisms acting
on the \hi\ and stellar content of massive galaxies 
{\it in the group environment} and from a statistical point of view.

\subsection{Quantifying the suppression of \hi\ gas}

If environmental mechanisms play an important role in removing cold
gas from galaxies, it is reasonable to expect that the \hi\ content of
the affected galaxies will be lower than that of similar (in terms of
structural and star formation properties), but
unperturbed, systems. This idea is behind the definition of the
classic ``\hi\ deficiency'' parameter \citep{haynes84}, which has been successfully
used to demonstrate that galaxies in the densest environments have
their \hi\ gas content largely reduced, most likely by ram pressure
stripping by the dense intracluster medium 
\citep{gh85,solanes01,chung07,vollmer09,luca11}.

As mentioned in Section~\ref{s_gfplane}, the gas fraction plane
is an excellent tool to investigate environmental effects, and the
distance from the best fit relation has been shown to be
equivalent to the \hi\ deficiency for galaxies in the Virgo cluster
\citep{luca11}. Indeed, the plane is a reformulation of the
\hi\ deficiency relation in terms of quantities (stellar mass surface
density and \nuvr\ color, which is a proxy for specific star formation
rate) that have a more immediate physical interpretation (compared to
morphological classification and optical diameter) and are more easily
applicable to large, modern data sets.
However, we show below that GASS does not probe the 
\hi-deficient regime, hence the gas fraction plane is of limited use to
find evidence for gas suppression {\it within our own sample}.

\begin{figure*}
\includegraphics[width=15cm]{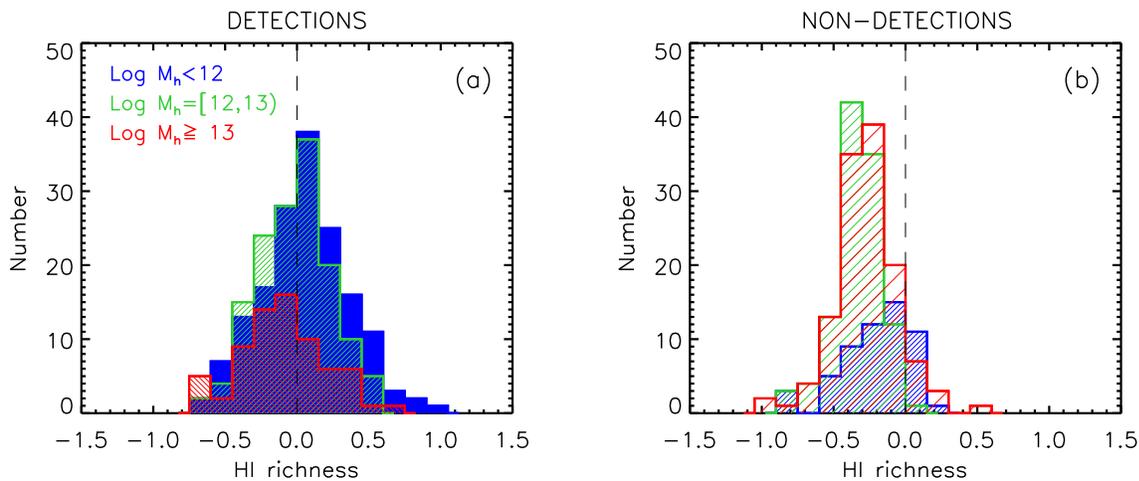}
\caption{$(a)$ Distribution of \hi\ richness, \ie\ the difference between
measured and expected gas fractions, for \hi-detected galaxies. The sample is divided into 
three bins of halo mass, as indicated in the top left corner. $(b)$ Same distributions for galaxies 
that were not detected in \hi\ (plotted at their upper limits), together with detections below 
the gas fraction limit of GASS. The colors correspond to the same halo mass bins indicated 
on the top left corner of panel (a).}
\label{HIexcess}
\end{figure*}

We measured the ``\hi\ richness'' parameter for our galaxies, defined as the
difference between the logarithms of the measured gas fraction and
that predicted by the relation in Figure~\ref{plane}a. \hi-poor
galaxies have smaller gas fractions than predicted, \ie\ a negative \hi\ richness.
Figure~\ref{HIexcess}a shows the distribution of the \hi\ richness parameter
for \hi-detected galaxies in three bins of halo mass, as indicated on the top left;
panel (b) shows the same histograms for the non-detections. These distributions
clearly illustrate that the non-detections pile up at \hi\ richness between $-0.5$ and $-0.2$, 
but they start to be important already in the ``\hi-normal'' regime, \ie\ near
\hi\ richness of zero.
Keeping in mind that the scatter of the plane is 0.3 dex, only
galaxies with gas fractions that deviate from the predicted values by
at least that amount (or more conservatively 0.5 dex, as usually
assumed for the \hi\ deficiency parameter) should be called
\hi-deficient (or \hi-excess) systems. Thus, because our survey
gas fraction limit is so close to the start of the \hi-deficient
regime, it turns out that the plane is much better suited to characterize
the \hi-excess systems in the GASS redshift interval than the \hi-poor
ones (since for the latter we only have upper limits).

The sample used to define the gas fraction plane is indeed representative of 
unperturbed systems, because it does not include the \hi\ non-detections,
which are the galaxies affected by the environment. We checked this by 
computing the gas fraction plane using only \hi\ detections in the 
\Mh $< 10^{12}$ \Msun\ bin, which gives a solution that is indistinguishable 
from that in Figure~\ref{plane}a. The highest halo mass bin, \Mh $\geq 10^{13}$ \Msun,
includes only 70 detections, and although the corresponding gas fraction plane
is slightly offset towards lower \Mhi/\Mst\ with respect to the ``undisturbed''
one, the difference is statistically not significant (the mean difference
between the two solutions is 0.17 dex, with a standard deviation of 0.08 dex,
and the scatters of the planes are both 0.3 dex; see Fig.~\ref{HIexcess}a).

Because we cannot quantify the degree of \hi\ removal in
individual \hi-deficient systems at the distances probed by GASS, and 
also our statistics become limited when we start binning galaxies by
stellar mass and environment, we do not attempt to compute the
average gas fraction scaling relations presented in 
Figure~\ref{scalings} in bins of environmental density (see however
Section 7). This approach
was adopted by \citet{luca11} to compare Virgo cluster and \hi-normal
galaxies, and was successful because the more nearby Herschel
Reference Survey \citep[HRS;][]{hrs} sample 
includes \hi\ detections and more stringent upper limits in the
\hi-deficient regime.

Instead, as already done by \citet{coldgass3} for our sample, we adopt
the gas fraction threshold of GASS as the
nominal division between \hi-normal and \hi-deficient systems,
and look for trends in the \hi\ detection fractions as a function of
galaxy properties and environment. As discussed above, this is
entirely justified by the fact that the detection limit of GASS
roughly corresponds to the gas fraction separating \hi-normal from
\hi-deficient massive galaxies. In order to compute meaningful
detection fractions we excluded from our sample the objects for which
confusion within the Arecibo beam is certain (15\% of the
\hi\ detections, indicated by blue stars in Fig.~\ref{plane}; these
galaxies were not included in Fig.~\ref{HIexcess}). Also, as already
noted, the few \hi\ detections with gas fraction below the GASS limit
(dark green squares in Fig.~\ref{plane}) are effectively \hi-poor
systems, and thus are counted as non-detections.

\subsection{Suppression of \hi\ gas in the group environment}

In this section we investigate the relation between gas content and
other galaxy properties in different environments, looking for
possible evidence of gas removal at the highest densities. We use dark
matter halo masses as our environmental estimator and, for the
reasons explained above, we resort to using detection fractions to
characterize the average gas content in a given bin of, \eg, stellar
and halo mass.

\begin{figure*}
\includegraphics[width=16cm]{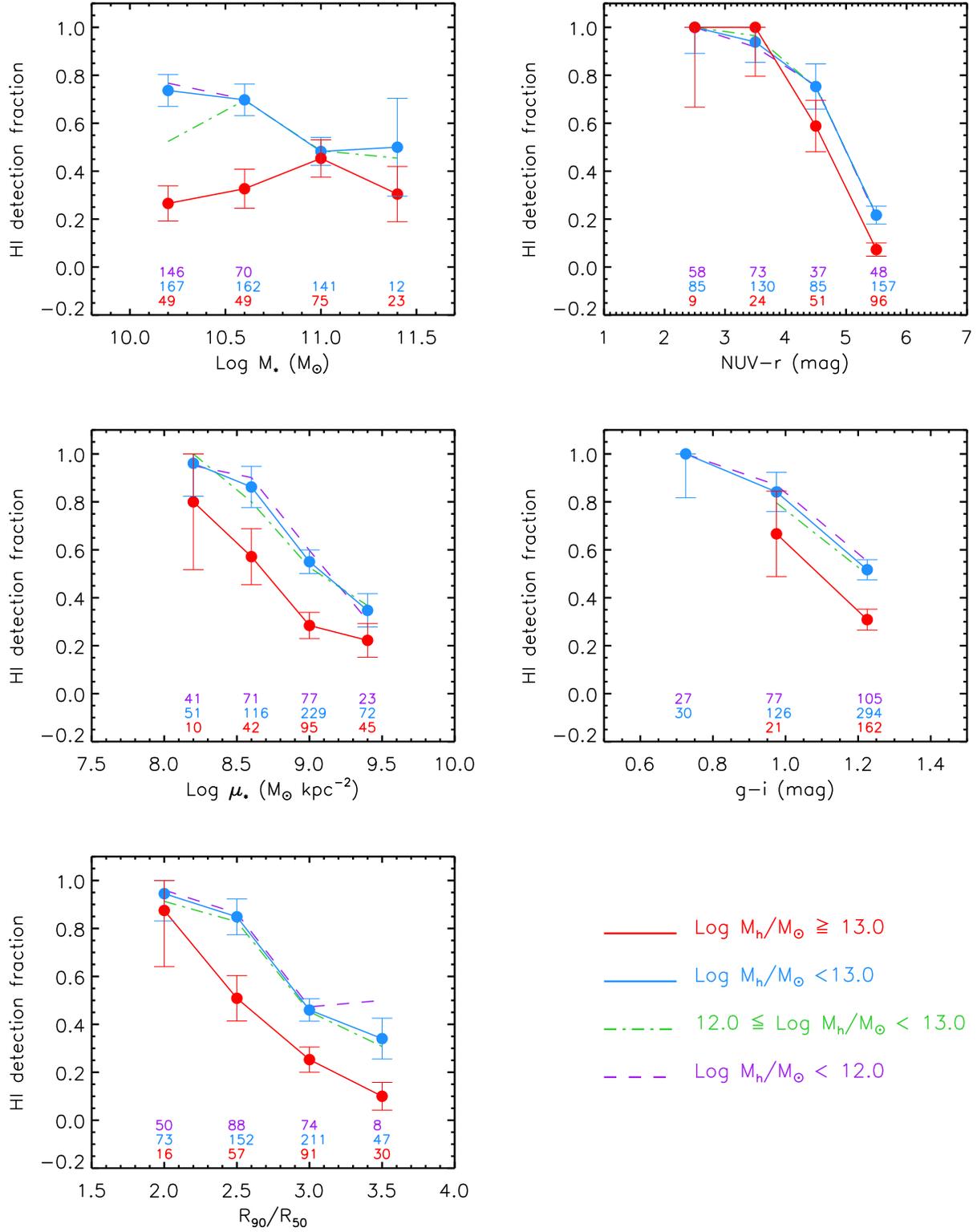}
\caption{\hi\ detection fraction of GASS galaxies plotted as a
function of stellar mass, stellar mass surface density and
concentration index in the first column, and \nuvr, \gi\ colors in the
second one. The data in each panel are divided into two bins of halo
mass, below and above $10^{13}$ \Msun\ (blue and red,
respectively), as indicated in the bottom right corner of the figure.
Large circles are average detection fractions, and the numbers in each panel
indicate the total number of galaxies in each bin (only bins with
\ntot $\geq 5$ are shown); errorbars are
Poissonian (truncated at detection fraction of 1 if necessary). We
also show the results for a finer division of the
lowest halo mass interval, \ie\ Log \Mh/\Msun $<12$ (dashed purple line)
and  $12 \leq$ Log \Mh/\Msun $<13$ (dot-dashed green line). Notice
that halos with \Mh $<10^{12}$ \Msun\ are populated only by galaxies
in the lowest two stellar mass bins.}
\label{detfr_halo}
\end{figure*}

\begin{figure*}
\includegraphics[width=16cm]{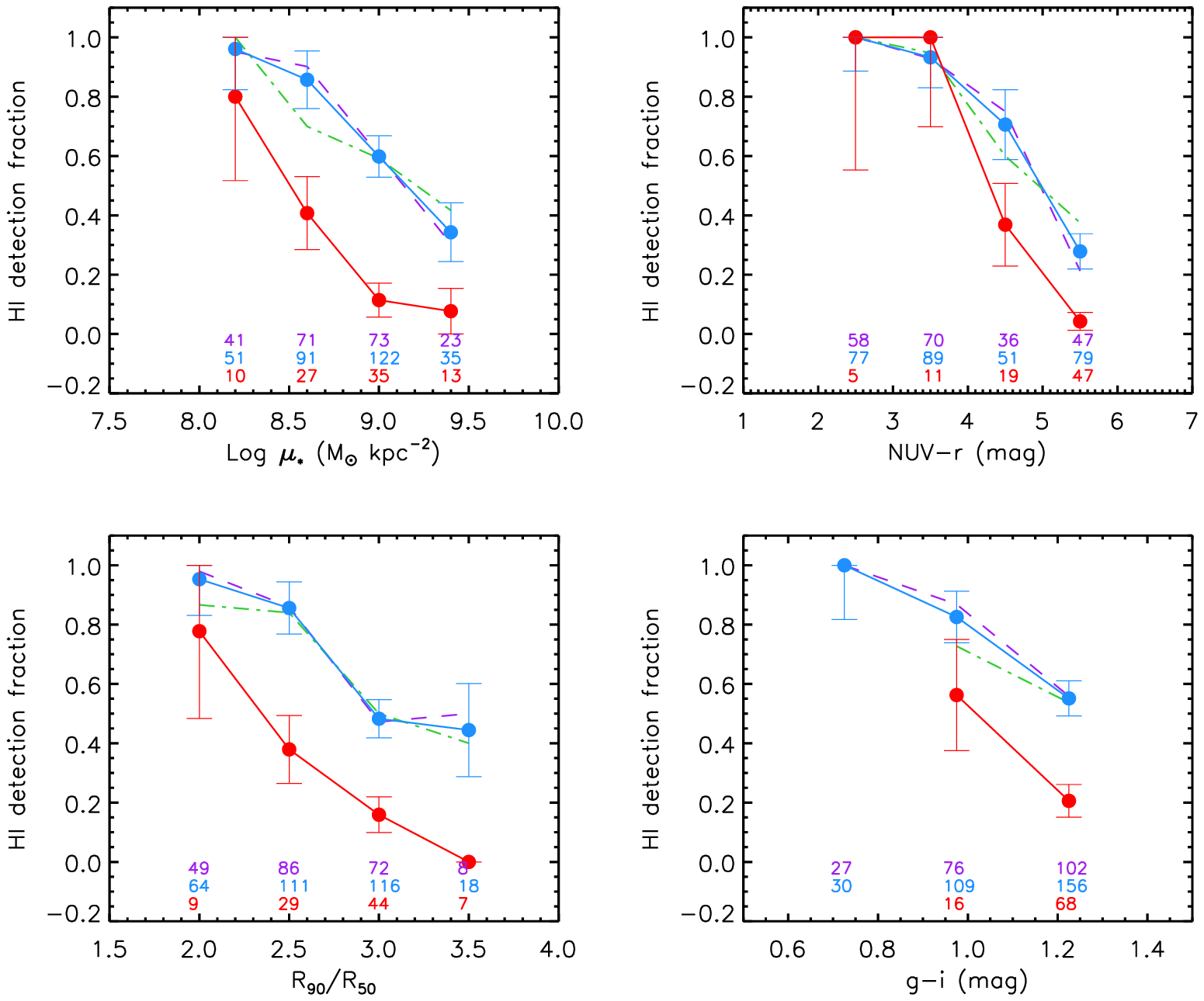}
\caption{\hi\ detection fraction plotted as a
function of stellar mass surface density, concentration index, \nuvr,
and \gi\ colors for the subset of galaxies with stellar mass Log \Mst/\Msun $< 10.75$.
Symbols and colors are the same as in Figure~\ref{detfr_halo}.}
\label{detfr_halo_lowMst}
\end{figure*}

Figure~\ref{detfr_halo} shows how the average \hi\ detection fraction,
\ie\ the ratio of detections to total in each bin, \detfr,
changes as a function of stellar mass, stellar mass surface density and
concentration index in the first column, and \nuvr, \gi\ colors in the
second one. Blue and red circles indicate galaxies
that inhabit dark matter halos with masses below and above 
$10^{13}$ \Msun, respectively.
We initially divided our sample into the same three bins
of halo mass used for Figure~\ref{HIexcess}, which contain similar
numbers of galaxies (see also Fig.~\ref{env_hist}). We indicate the
average detection fractions in the two lowest halo mass intervals,
Log \Mh/\Msun $<12$ and  $12 \leq$ Log \Mh/\Msun $<13$, with
dashed purple and dot-dashed green lines, respectively.
As can be seen, there is no significant difference between these two
halo mass bins in any of these plots (the only apparent exception
would be the first stellar mass bin in the top left panel, but notice
that the green data point is based on 20 objects only), so we combined
them to increase statistics.

The top left panel of Figure~\ref{detfr_halo} is the main result of this work, and
clearly shows that the \hi\ content of massive galaxies that live
in dark matter halos with \Mh $\gtrsim 10^{13}$ \Msun\ is
significantly reduced compared to that of galaxies {\it with the same
stellar mass}, at least below \Mst \about $10^{11}$ \Msun. We do not see a difference
at larger stellar mass, which seems to suggest that the environment
has no detectable effect on the most massive galaxies in our
sample. We will come back to this point later. 
Because GASS does not contain any very rich group or clusters (and
indeed 2/3 of the halos in our highest density bin have masses between 
$10^{13}$ and $10^{14}$ \Msun; see also Fig.~\ref{env_hist}b), our result
implies that {\it the suppression of} \hi\ {\it is modulated by the
environment even at the intermediate densities probed by our sample.}

The other two panels in the first column of Figure~\ref{detfr_halo} show that the
suppression of \hi\ gas in the most massive halos in our sample can be
seen also at fixed stellar mass surface density and concentration
index, both proxies of stellar morphology (higher values of \must\ and
\cindx\ correspond to bulge-dominated systems).

\begin{figure*}
\includegraphics[width=16cm]{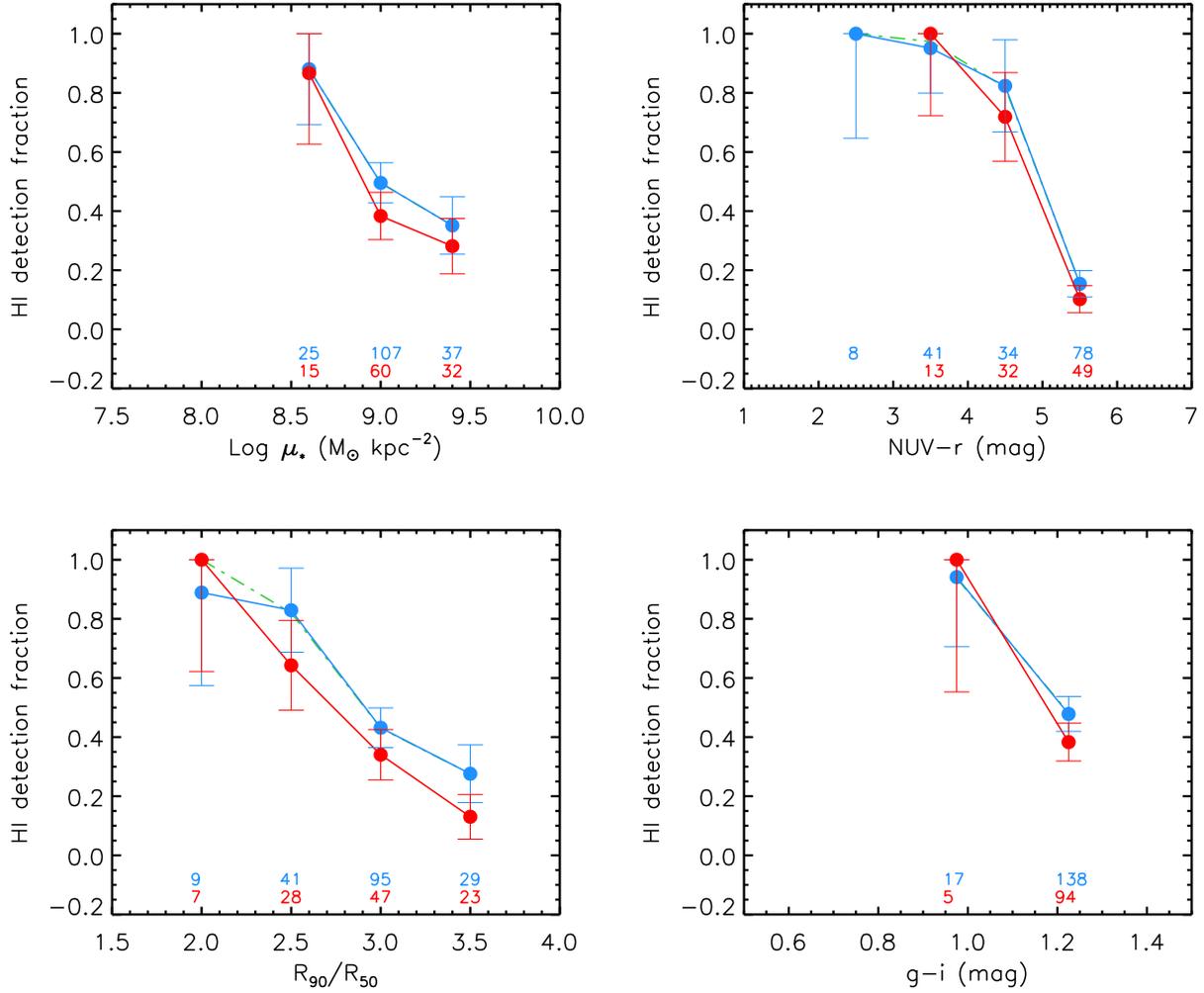}
\caption{Same as Figure~\ref{detfr_halo_lowMst} for galaxies with
larger stellar mass (Log \Mst/\Msun $\geq 10.75$).}
\label{detfr_halo_highMst}
\end{figure*}

\begin{figure*}
\includegraphics[width=16cm]{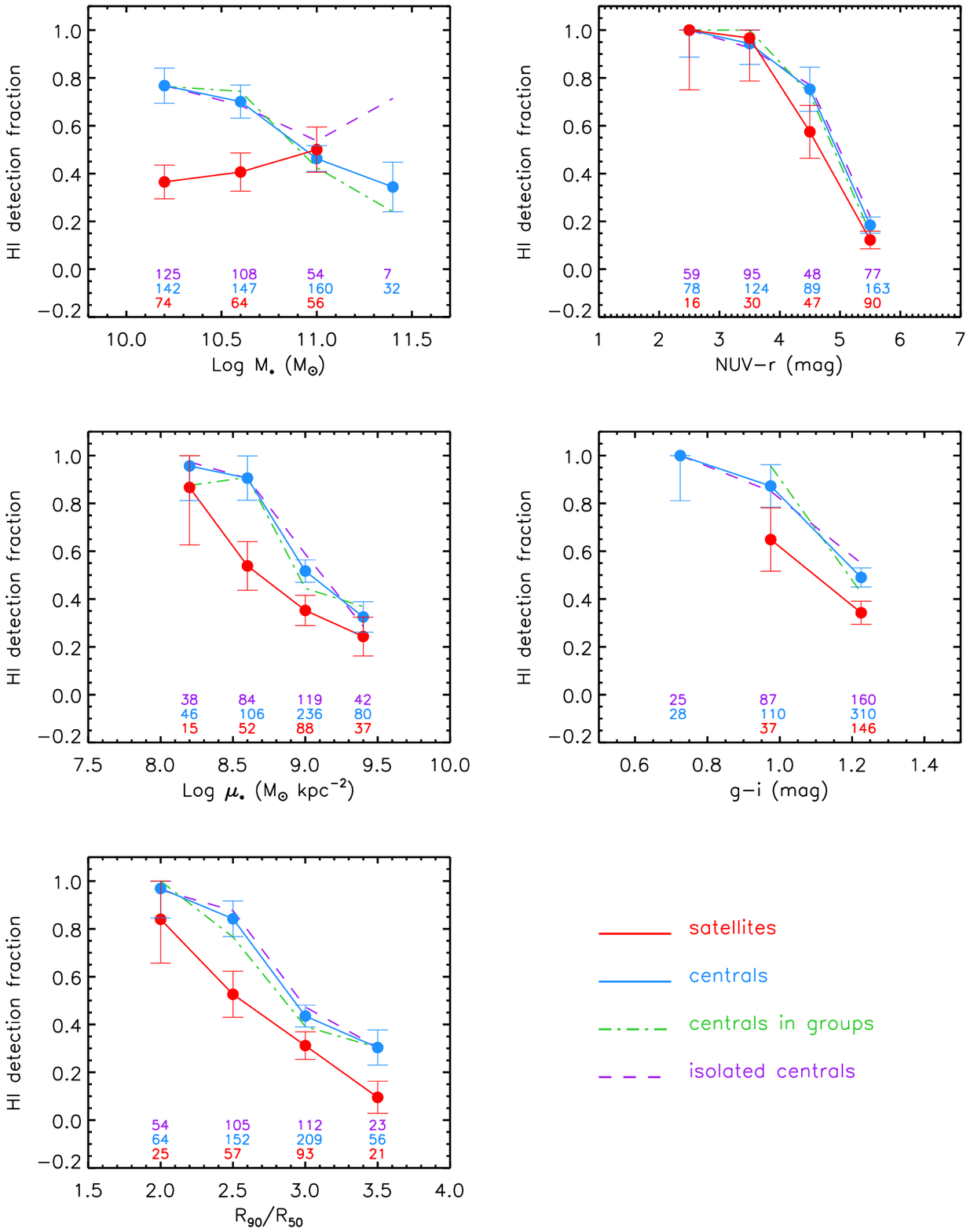}
\caption{The \hi\ detection fraction of GASS galaxies is plotted here
as a function of the same quantities seen in Figure~\ref{detfr_halo},
but now the data are divided into centrals (blue) and satellites
(red). The purple dashed and green dot-dashed lines indicate central
galaxies in isolation and in groups, respectively.}
\label{detfr_centr}
\end{figure*}

The plots on the right column of Figure~\ref{detfr_halo} compare the
detection fractions in different environments at fixed galaxy color.
Interestingly, we find that galaxies in more massive halos 
have lower gas content only for \nuvr\ colors redder than \about 4 mag. 
In the stellar mass range probed by GASS, this color corresponds 
to the red edge of the blue cloud and the start of the green valley 
\citep{wyder07}, suggesting that a fraction of our gas-poor systems 
have not yet completely stopped forming stars. 
The presence of gas-poor, but still star-forming, galaxies may 
indicate that the timescale of the gas removal is significantly 
shorter than the timescale necessary for the \nuvr\ color to reach
values typical of the red sequence galaxies, \ie\ \nuvr \about 5.5 mag
(\about 1 Gyr, see also Fig.~4 in \citealt{luca11}).

Less enlightening is the variation of \hi\ detection fraction with 
\gi\ color. Although we find that, at fixed \gi\ color, galaxies in 
high mass halos have significantly lower detection fractions, this result 
does not provide any additional insights into the physical process at play. 
Indeed, massive galaxies generally lie on the optical red sequence 
regardless of their current star formation activity 
\citep{wyder07,luca12_redsp}, thus their
optical colors are saturated --- they cannot significantly redden
following further quenching of the star formation.

We look in more detail at the properties of the lower stellar mass
galaxies, for which we see a clear difference of gas content above and
below \Mh \about $10^{13}$ \Msun, in Figure~\ref{detfr_halo_lowMst}.
Here the detection fraction is shown as a function of stellar mass 
surface density, concentration index, \nuvr, and \gi\ colors for the 
subset of galaxies with \Mst $< 10^{10.75}$ \Msun. For comparison, the
same plots are presented in Figure~\ref{detfr_halo_highMst} for the
galaxies with stellar mass above that limit. As expected, the offsets
seen in Figure~\ref{detfr_halo} become larger when we restrict
the sample to the lower stellar mass bin. This is particularly
interesting in the case of the \nuvr, since it slightly reinforces our
timescale argument. Overall, the larger differences shown in
Figure~\ref{detfr_halo_lowMst} are simply due to the exclusion of the 
most massive galaxies, which have lower gas fractions (see
Fig.~\ref{scalings}). With regard to the
galaxies with stellar mass \Mst $\geq 10^{10.75}$ \Msun, we caution
the reader that the median galaxy is a non-detection, hence we
cannot conclude that the environment is not acting on the gas
reservoir of those systems -- our survey might simply not be sensitive
enough to detect environmental effects on these already gas-poor galaxies.

The trends in detection fraction observed when we divide the sample
according to halo mass are present also when we describe the environment
in terms of central and satellite galaxies. Figure~\ref{detfr_centr}
repeats the panels of Figure~\ref{detfr_halo}, but now blue and red
circles represent central and satellite galaxies,
respectively. Purple dashed and green dot-dashed lines indicate central
galaxies in isolation and in groups, respectively. There is no
significant difference between the two classes of central galaxies
and, at fixed stellar mass (at least below \about $10^{11}$ \Msun), satellite
galaxies have lower gas content on average than centrals. This is
completely consistent with the result shown in the corresponding panel
of Figure~\ref{detfr_halo}, as expected from the fact that central
galaxies in this stellar mass interval are mostly isolated
(see Fig.~\ref{mst_mh}). Overall, the offsets in detection fraction
are slightly smaller when we divide the sample into central and satellites
rather than by halo mass (mostly because satellite galaxies are found
at all halo masses, not only in halos with \Mh $>10^{13}$ \Msun), but
they are still significant.

It would be very interesting to know whether the observed decrease of
\hi\ content is primarily dependent on the dark matter halo mass or on
the nature of the galaxy as central vs. satellite. This is because
there could be physically distinct processes that link \hi\ content
separately to these two different environmental descriptors 
\citep[\eg][]{weinmann06a}. Unfortunately, our data 
do not allow us to disentangle between the two scenarios. As can be
seen by simply drawing a horizontal line at Log \Mh/\Msun $=13$ in 
Figure~\ref{mst_mh}, there are almost no central galaxies above that
threshold and there are only very few satellites below. Therefore,
although splitting the sample by halo mass is not the same as splitting by
centrals vs. satellites, once we bin the galaxies to reach sufficient
statistics the two classifications become almost the same, and the
issue ends up being just a semantic one.

\begin{figure}
\includegraphics[width=8cm]{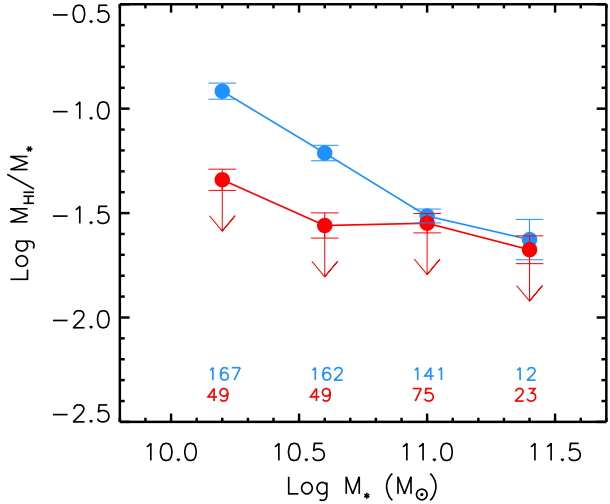}
\caption{Averages of \hi\ gas fraction logarithms versus stellar mass.
The data are divided into two bins of halo mass, above and below 
$10^{13}$ \Msun\ (red and blue, respectively). Downward arrows
indicate upper limits (because the corresponding bins are dominated by
\hi\ non-detections). The numbers at the bottom indicate the total
number of galaxies in each bin (only bins with \ntot $\geq 5$ are
shown); errorbars are errors on the mean.}
\label{gf_Mst_halo}
\end{figure}

\section{Discussion and conclusions}

In this work we have used the full GASS data set, which includes
\hi\ measurements for \about 800 galaxies with stellar masses 
$10 < {\rm Log} (M_\star/M_\odot) < 11.5$
and redshift $0.025 < z < 0.05$, to study how the gas content of
massive systems depends on environment {\it at fixed stellar mass}. 
We characterized the environment of GASS galaxies by their dark matter
halo mass, obtained from the SDSS group catalog of 
\citet[][updated to SDSS DR7]{yang07} using the abundance
matching technique.

The key new result of our analysis is that we obtained clear evidence for
suppression of \hi\ gas at fixed stellar mass (at least 
below \Mst \about $10^{11}$ \Msun) for galaxies that are located in
groups with halo masses \Mh $\gtrsim 10^{13}$ \Msun.
The effect is seen also at fixed stellar morphology (\ie, \must\ and
\cindx), and when we divide our sample according to central/satellite
classification.
As shown in Figure~\ref{env_hist}, our most massive halo bin is
dominated by systems with \Mh\ between $10^{13}$ and $10^{14}$ \Msun. In the
SDSS group catalog, such halos include up to \about 60 members (20 on
average), whereas smaller halos include up to 10 members (2 on
average). Thus, the environment where we detect a decrease of \hi\ gas
content in massive galaxies is that of moderately rich groups, and we
are certainly not probing the cluster regime.
 
We attempt to quantify the amount of gas depletion for our sample in Figure~\ref{gf_Mst_halo}.
We computed average gas fractions in bins of stellar and halo mass,
including the non-detections at their upper limits. As in 
Figures~\ref{detfr_halo}-\ref{detfr_halo_highMst}, blue and red
lines indicate dark matter halos with masses below and above $10^{13}$
\Msun, respectively. The result is 
qualitatively consistent with what shown in the top left panel of Figure~\ref{detfr_halo} for
the average detection fractions: at fixed stellar mass (at least below \about
$10^{11}$ \Msun), the \hi\ content of galaxies in more massive halos
is systematically lower. In the first two stellar mass bins, the
difference of \hi\ gas fractions between galaxies in halos with 
masses below and above $10^{13}$ \Msun\ is \about 0.4 dex (linear gas
fractions drop from 12\% to 5\% in the first \Mst\ bin, and from 6\%
to 3\% in the second one). As indicated by the red arrows, the
average gas fractions for the \Mh $\geq 10^{13}$ \Msun\ bins (and
those for \Mst $\geq 10^{11}$ \Msun\ regardless of halo mass) are dominated by 
non-detections, and thus must be considered upper limits. This gives
us a lower limit on the typical amount of \hi\ suppression in
groups, which is at least a factor of two compared to galaxies in smaller
halos, but prevents us from a more precise quantification. This is the
reason why we decided to carry out our analysis in terms of detection
fractions instead of gas fractions.

As expected, the decrease of \hi\ content measured in the group
environment for our sample, 0.4 dex, is smaller than what observed in
higher density regions, such as rich galaxy clusters. For instance,
\hi-deficient Virgo members with stellar masses \about
$10^{10}-10^{10.7}$~\Msun\ have gas fractions that are 0.8 dex smaller
than \hi-normal galaxies in the HRS 
\citep[see table 1 in][]{luca11}. Galaxies in more massive clusters
such as Coma have more extreme levels of \hi\ deficiency \citep{solanes01}.

\begin{figure}
\includegraphics[width=8cm]{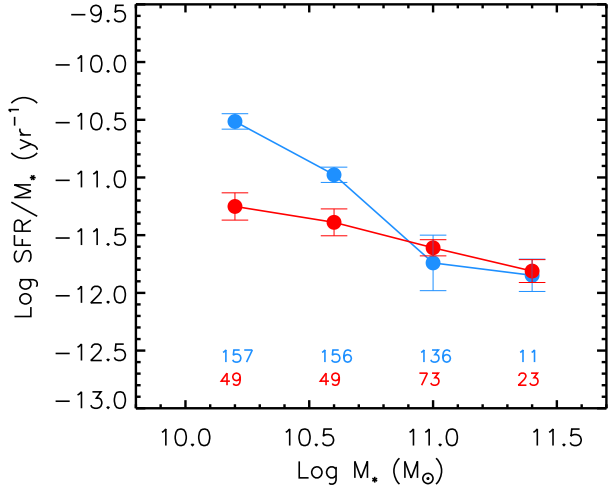}
\caption{Average specific star formation rates are plotted as a
function of stellar mass for two bins of halo mass (symbols and
colors as in Fig.~\ref{gf_Mst_halo}).}
\label{ssfr_Mst_halo}
\end{figure}

It is interesting to determine whether the star formation
properties of the galaxies for which \hi\ has been reduced are
affected as well. Figure~\ref{ssfr_Mst_halo} shows the running
averages of the specific star formation rates versus stellar mass for
our sample, binned by halo mass as in Figure~\ref{gf_Mst_halo}. The
star formation rates were computed from our NUV photometry as in
\citet{gass2}. As for the gas, we see a quenching of the star formation
in the group environment (at least for galaxies with stellar mass less than
\about $10^{11}$ \Msun). This is in qualitative agreement with optical
studies, which established that the star formation properties of
galaxies are affected by the environment well before reaching the
high-density regimes that are typical of clusters
\citep[\eg][]{lewis02,gomez03}. 
A detailed comparison with such studies is difficult, as sample
selections and environmental descriptors vary widely, and we
specifically targeted only massive galaxies.

\begin{figure}
\includegraphics[width=8cm]{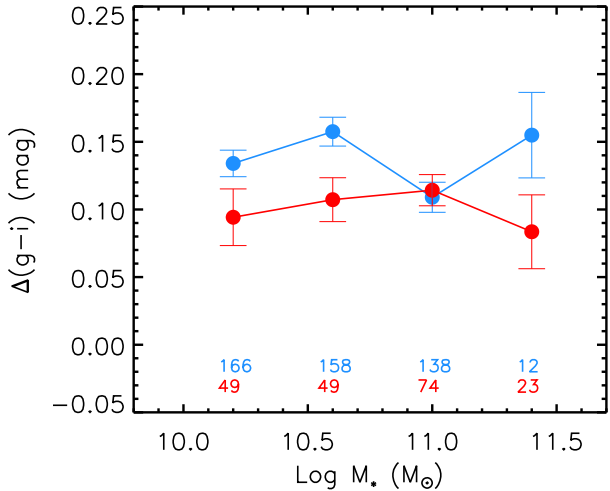}
\caption{Average \gi\ color gradients, defined as the difference
between inner and outer \gi\ colors, versus stellar mass for two
bins of halo mass (symbols and colors as in Figures~\ref{gf_Mst_halo} and
\ref{ssfr_Mst_halo}).}
\label{gi_diff_Mst_halo}
\end{figure}

We can think of two main scenarios to explain the observed suppression
of \hi\ content in group galaxies: direct removal of \hi\ from the disk
and starvation \citep{larson80}. In the first case, the \hi\ is directly
affected and removed from the galaxy disk by one or more environmental
mechanisms (\eg, ram pressure or gravitational interactions).
In the second case, the lower \hi\ mass fraction in the more massive
halos (and in satellites vs. centrals) is due to the group environment
disrupting the accretion of the infalling, pristine gas, which, if allowed
to reach the galaxy disk, would subsequently replenish its \hi\
reservoir.
However, it seems unlikely that starvation alone could explain both
the \hi\ suppression and the difference of gas content at fixed
specific star formation rate seen in our data. If the supply of infalling gas is
stopped and no other external mechanisms are at play, then the \hi\ in
the galaxy will be consumed by star formation, and the two quantities
should track each other and decrease on the same timescale
\citep{boselli06,luca11}. Instead, Figure~\ref{detfr_halo_lowMst} shows that, at
fixed \nuvr\ color (\ie, at fixed specific star formation rate),
the \hi\ content of galaxies in more massive halos is
systematically lower, at least in objects with stellar masses less
than $10^{11}$ \Msun. This supports a scenario in which an environmental
mechanism acting directly on the cold gas reservoir is needed to
explain our findings. We will assume that this is the case in the remainder
of this section.

Without detailed information on the distribution and kinematics of the \hi\ gas
we cannot determine which environmental process is responsible
for the \hi\ removal, but we can try to establish if it acts
outside-in by looking at the color gradients of our galaxies. 
Indeed, \citet{luca12} have recently shown that the extent of the
star-forming disk and the shape of the color gradients are tightly
related to the amount of \hi\ gas. Using \gi\ color gradients of
massive galaxies extracted from the GASS parent sample, 
\citet{gass3} showed that more \hi-rich
systems are bluer on the outside relative to the inside compared to
control samples matched in stellar mass and redshift. We use the same
quantity adopted by \citet{gass3}, but with opposite sign, and define
$\Delta(g-i)$ as the difference between inner and outer \gi\ colors
(inner and outer regions are enclosed by \rhalf\ and 2.5 times the
Kron radius, both determined from \rband\ photometry,
respectively). Therefore $\Delta(g-i)$ is typically positive for disk
galaxies (especially the bulge-dominated ones), because their outer
regions are bluer than their inner regions.
We plot the average \gi\ color gradients versus stellar mass and in our
two usual halo bins in Figure~\ref{gi_diff_Mst_halo}. There is
tentative evidence that galaxies in the stellar mass interval of
interest ($\lesssim 10^{11}$ \Msun) have smaller values of $\Delta(g-i)$ when they
are located in more massive halos --- in other words their color
gradients are {\it flatter}. Because their specific star formation
rates are smaller (\ie, their global \gi\ colors are redder),
this implies that their outer regions have become
redder (as opposed to their central parts bluer), compared to those of
galaxies with the same stellar mass but found in smaller halos.
This is expected from the fact that most of the \hi\ gas in a galaxy
is typically found beyond \rhalf, and supports an outside-in
suppression (without any strong enhancement in the center) of both gas
and star formation in groups.

From the evidence presented by our data, we conclude that \hi\ gas is
removed from massive galaxies in the group environment, and that the
process responsible for this quenches their star formation as well,
most likely in the outer regions of the galaxy. Although we clearly
observe the \hi\ suppression only in galaxies with
stellar masses less than \about $10^{11}$ \Msun, we cannot exclude
that environmental effects are at work also in more massive systems,
which are already gas poor. This is because at high stellar mass the
average GASS galaxy is a non-detection, hence we are not able to detect a
possible \hi\ decrease with respect to similar objects in
smaller dark matter halos.

As discussed in the previous section, the 
difference of detection fractions at fixed \nuvr\ color between high
and low mass halos might indicate 
that the suppression of the gas takes place on timescales of \about 1 Gyr
or shorter. This would be in qualitative agreement with the 
cosmological hydrodynamical simulations of \citet{dave13}, which
suggest that the process that removes \hi\ from satellite galaxies
acts quickly compared to the infall timescale into the halo (several
Gyrs). All this points to a pre-processing of the gas (and star
formation) in the group environment. Both ram pressure stripping and
tidal interactions might be responsible for this quenching, but the fact
that the mechanism seems to truncate the star formation outside-in
might favor ram pressure. It is currently unclear if ram pressure
stripping can significantly affect the interstellar medium of galaxies
outside the rich cluster environment, where hot X-ray-emitting gas is
not present, but there is some evidence that this might be the case
\citep[\eg][]{scott12,freeland11}.

Very interestingly, \citet{fabello3} came to a similar conclusion with
a completely different approach. These authors determined the average gas content of
massive galaxies by cross-correlating the GASS parent sample with
ALFALFA, and stacking the \hi\ spectra (mostly non-detections). They
binned the galaxies by stellar mass and local density, estimated from
the number of neighbors with \Mst $\geq 10^{9.5}$ \Msun\ within 1 Mpc
and $\pm 500$ \kms, and compared their results with predictions of
semi-analytic models \citep{guo11}. For galaxies
with \Mst $\leq 10^{10.5}$ \Msun\ (where they are not limited by small
number statistics), the decline in average gas fraction with local density is
stronger than the decline in mean global and central specific star
formation rates. This ordering is not reproduced by the semi-analytic
models, which do not include stripping of the cold interstellar
medium, and suggests that ram pressure is able to remove
atomic gas from the outer disks of galaxies {\it in the group
environment} probed by GASS. Furthermore, \citet{fabello3} used mock
catalogs generated from the semi-analytic models to show that galaxies with 
$10 < $ log \Mst/\Msun $< 10.5$ and local density parameter $N>7$, for
which the strong decline in \hi\ content is seen, are found in dark
matter halos with masses in the range of $10^{13}-10^{14}$ \Msun, in
agreement with what we determined more directly in this work.

Although it is well known that the star formation of galaxies is
affected by the environment well before reaching the highest densities
typical of clusters, to our knowledge this is the first time
that environmental effects have been proved to remove \hi\ gas in
groups in a statistical sense and from an observational point of view.
Our data indicate that, at fixed stellar mass, the gas fraction of
galaxies with stellar mass between $10^{10}$ and $10^{11}$ \Msun\ drops
by at least 50\% in dark matter halos with \Mh \about
$10^{13}-10^{14}$ \Msun. The removal of gas in groups most likely drives 
the observed quenching of the star formation in these systems, and
although not conclusive, we offered some evidence in support of a
hydrodynamical process like ram pressure stripping behind this effect.
This is extremely important for our understanding of the physical
processes that transform galaxies from blue, star-forming to red and
passively evolving, and suggests a key role for the pre-processing in
groups. Indeed, hydrodynamical processes are usually considered not
to be important in groups, and simulations do not include them (for
instance, in the \citealt{guo11} models, tidal and ram-pressure forces
only remove hot gas from the halos of infalling satellites, and do not
act on the cold gas).

Progress in this field requires not only better
statistics and spatial resolution, but also sensitivity to low
levels of gas content, which can be achieved only with large apertures
and/or long integrations. GASS has the unique advantage of combining a
stellar mass selection over a large volume (100-200 Mpc) with a low
gas fraction limit, which allowed us to detect galaxies with
\Mhi/\Mst\ down to a few per cent. In order to reach these gas
fractions, we observed our targets up to 90 minutes on-source with the
largest collecting area currently available. Restricting the survey to
lower redshifts would decrease the telescope time, but at the price of
increasing cosmic variance.
All-sky \hi-blind surveys planned with the Australian
Square Kilometer Array (SKA) Pathfinder \citep[ASKAP;][]{askap} and the upgraded 
Westerbork Synthesis Radio Telescope \citep[APERTIF;][]{apertif}, will
provide larger samples and much better spatial resolution. The large
volumes surveyed will compensate for the modest sensitivity, which
will be comparable to that of ALFALFA, definitely allowing a step
further in this field. Furthermore, stacking is a promising,
complementary technique to extend the results presented in this work. 
However, a sensitive \hi\ survey able to detect
galaxies with small gas fractions over a comparable volume to GASS and
across a wide range of environments might have to wait for the full SKA.

\section*{Acknowledgments}

We thank the anonymous referee for useful suggestions. 

BC is the recipient of an Australian Research Council Future 
Fellowship (FT120100660). APC acknowledges a National Natural Science Foundation of China
international cooperation and exchange grant, no. 11250110509.
The research leading to these results has received funding from the
European Community's Seventh Framework Programme (/FP7/2007-2013/)
under grant agreement No 229517.

This research has made use of the NASA/IPAC Extragalactic Database
(NED) which is operated by the Jet Propulsion Laboratory, California
Institute of Technology, under contract with the National Aeronautics
and Space Administration.

The Arecibo Observatory is operated by SRI International under a
cooperative agreement with the National Science Foundation
(AST-1100968), and in alliance with Ana G. M{\'e}ndez-Universidad
Metropolitana, and the Universities Space Research Association.

GALEX (Galaxy Evolution Explorer) is a NASA Small Explorer, launched
in April 2003. We gratefully acknowledge NASA's support for
construction, operation, and science analysis for the GALEX mission,
developed in cooperation with the Centre National d'Etudes Spatiales
(CNES) of France and the Korean Ministry of Science and Technology. 

Funding for the SDSS and SDSS-II has been provided by the Alfred
P. Sloan Foundation, the Participating Institutions, the National
Science Foundation, the U.S. Department of Energy, the National
Aeronautics and Space Administration, the Japanese Monbukagakusho, the
Max Planck Society, and the Higher Education Funding Council for
England. The SDSS Web Site is http://www.sdss.org/.

The SDSS is managed by the Astrophysical Research Consortium for the
Participating Institutions. The Participating Institutions are the
American Museum of Natural History, Astrophysical Institute Potsdam,
University of Basel, University of Cambridge, Case Western Reserve
University, University of Chicago, Drexel University, Fermilab, the
Institute for Advanced Study, the Japan Participation Group, Johns
Hopkins University, the Joint Institute for Nuclear Astrophysics, the
Kavli Institute for Particle Astrophysics and Cosmology, the Korean
Scientist Group, the Chinese Academy of Sciences (LAMOST), Los Alamos
National Laboratory, the Max-Planck-Institute for Astronomy (MPIA),
the Max-Planck-Institute for Astrophysics (MPA), New Mexico State
University, Ohio State University, University of Pittsburgh,
University of Portsmouth, Princeton University, the United States
Naval Observatory, and the University of Washington.

\bibliography{catinella}

\section*{Appendix A: Data Release}

We present here SDSS postage stamp images, Arecibo \hi-line spectra,
and catalogs of optical, UV and \hi\ parameters for the
250 galaxies included in this third and final data release. The
content and format of the tables are identical to the DR2 ones, and we
refer the reader to that paper for details. We only briefly summarize
their content below. Notes on individual objects
(marked with an asterisk in the last column of Tables~\ref{t_det} and
\ref{t_ndet}) are reported in Appendix~B.\\

\noindent
{\bf SDSS and GALEX data.}\\
Table~\ref{t_sdss} lists optical and UV quantities for the 250 GASS
galaxies, ordered by increasing right ascension:\\
Cols. (1) and (2): GASS and SDSS identifiers. \\
Col. (3): UGC \citep{ugc}, NGC \citep{ngc} or IC \citep{ic,ic2}
designation, or other name, typically from
the Catalog of Galaxies and Clusters of Galaxies \citep[CGCG;][]{cgcg}, 
or the Virgo Cluster Catalog \citep[VCC;][]{vcc}.\\
Col. (4): SDSS redshift, $z_{\rm SDSS}$. The typical uncertainty of
SDSS redshifts for this sample is 0.0002.\\
Col. (5): base-10 logarithm of the stellar mass, \Mst, in solar
units. Stellar masses are derived from SDSS photometry using the
methodology described in \cite{salim07} (a \citealt{chabrier03}
initial mass function is assumed).
Over our required stellar mass range, these values are
believed to be accurate to better than 30\%.\\
Col. (6): radius containing 50\% of the Petrosian flux in \zband, \Rinz,
in arcsec.\\
Cols. (7) and (8): radii containing 50\% and 90\% of the Petrosian
flux in \rband, $R_{50}$ and  $R_{90}$ respectively, in arcsec.\\
Col. (9): base-10 logarithm of the stellar mass surface density, \must, in
\Msun~kpc$^{-2}$. This quantity is defined as 
$\mu_\star = M_\star/(2 \pi R_{50,z}^2)$, with \Rinz\ in kpc units.\\
Col. (10): Galactic extinction in \rband, ext$_r$, in magnitudes, from SDSS.\\
Col. (11): \rband\ model magnitude from SDSS, $r$, corrected for Galactic extinction.\\
Col. (12): minor-to-major axial ratio from the exponential
fit in \rband, $(b/a)_r$, from SDSS.\\
Col. (13): inclination to the line-of-sight, in degrees (see
\citealt{gass6} for details).\\
Col. (14): \nuvr\ observed color from our reprocessed photometry,
corrected for Galactic extinction.\\
Col. (15): exposure time of GALEX NUV image, T$_{NUV}$, in seconds.\\
Col. (16): maximum on-source integration time, \tmax, required to
reach the limiting \hi\ mass fraction, in minutes (see \S~\ref{s_sample}).
Given the \hi\ mass limit and redshift of each galaxy, \tmax\ is
computed assuming a 5$\sigma$ signal with 300 \kms\ velocity width and
the instrumental parameters typical of our observations (\ie, gain
\about 10 K Jy\minusone\ and system temperature \about 28 K at 1370 MHz).\\

\noindent
{\bf \hi\ source catalogs.}\\
This data release includes 147 detections and 103 non-detections, for
which we provide upper limits below.

The measured \hi\ parameters for the detected
galaxies are listed in Table~\ref{t_det}, ordered by increasing right ascension:\\
Cols. (1) and (2): GASS and SDSS identifiers. \\
Col. (3): SDSS redshift, $z_{\rm SDSS}$. \\
Col. (4): on-source integration time of the Arecibo
observation, $T_{\rm on}$, in minutes. This number refers to
{\it on scans} that were actually combined, and does not account for
possible losses due to RFI excision (usually negligible). \\
Col. (5): velocity resolution of the final, smoothed spectrum in \kms. \\
Col. (6): redshift, $z$, measured from the \hi\ spectrum.
The error on the corresponding heliocentric velocity, $cz$, 
is half the error on the width, tabulated in the following column.\\
Col. (7): observed velocity width of the source line profile
in \kms, \whi, measured at the 50\% level of each peak. 
The error on the width is the sum in quadrature of the 
statistical and systematic uncertainties in \kms. Statistical errors
depend primarily on the signal-to-noise of the \hi\ spectrum, and are
obtained from the rms noise of the linear fits to the edges of the
\hi\ profile. Systematic errors depend on the subjective choice of the
\hi\ signal boundaries (see DR1 paper), and are negligible for most of
the galaxies in our sample (see also Appendix~B).\\
Col. (8): velocity width corrected for instrumental broadening
and cosmological redshift only, \whi$^c$, in \kms\ (see
\citealt{gass6} for details). No inclination or turbulent motion
corrections are applied.\\
Col. (9): observed, integrated \hi-line flux density in Jy \kms,
$F \equiv \int S~dv$, measured on the smoothed and baseline-subtracted
spectrum. The reported uncertainty is the sum in quadrature of the 
statistical and systematic errors (see col. 7).
The statistical errors are calculated according to equation 2 of S05
(which includes the contribution from uncertainties in the baseline fit).\\
Col. (10): rms noise of the observation in mJy, measured on the
signal- and RFI-free portion of the smoothed spectrum.\\
Col. (11): signal-to-noise ratio of the \hi\ spectrum, S/N,
estimated following \citet{saintonge07} and adapted to the velocity
resolution of the spectrum. 
This is the definition of S/N adopted by ALFALFA, which accounts for the
fact that for the same peak flux a broader spectrum has more signal.\\
Col. (12): base-10 logarithm of the \hi\ mass, \Mhi, in solar
units (see \citealt{gass6} for details).\\
Col. (13): base-10 logarithm of the \hi\ mass fraction, \Mhi/\Mst.\\
Col. (14): quality flag, Q (1=good, 2=marginal, 3=marginal and
confused, 5=confused). An asterisk indicates the presence of a note
for the source in Appendix~\ref{s_notes}.
Code 1 indicates reliable detections, with a S/N ratio of order of
6.5 or higher. Marginal detections have lower S/N, thus more uncertain
\hi\ parameters, but are still secure detections, with \hi\ redshift
consistent with the SDSS one. We flag galaxies as ``confused'' when
most of the \hi\ emission is believed to originate from another source
within the Arecibo beam. For some of the galaxies, the presence of
small companions within the beam might contaminate (but is unlikely to
dominate) the \hi\ signal -- this is just noted in Appendix~B.

Table~\ref{t_ndet} gives the derived \hi\ upper limits for the non-detections. 
Columns (1-4) and (5) are the same as columns (1-4) and (10) in Table~\ref{t_det},
respectively. Column (6) lists the upper limit on the \hi\ mass in
solar units, Log \Mhi$_{,lim}$, computed assuming a 5 $\sigma$ signal with 300 \kms\ 
velocity width, if the spectrum was smoothed to 150 \kms. Column (7)
gives the corresponding upper limit on the gas fraction, Log~\Mhi$_{,lim}$/\Mst.   
An asterisk in Column (8) indicates the presence of a note for the
galaxy in Appendix~B.\\

\begin{figure*}
\begin{center}
\includegraphics[width=16cm]{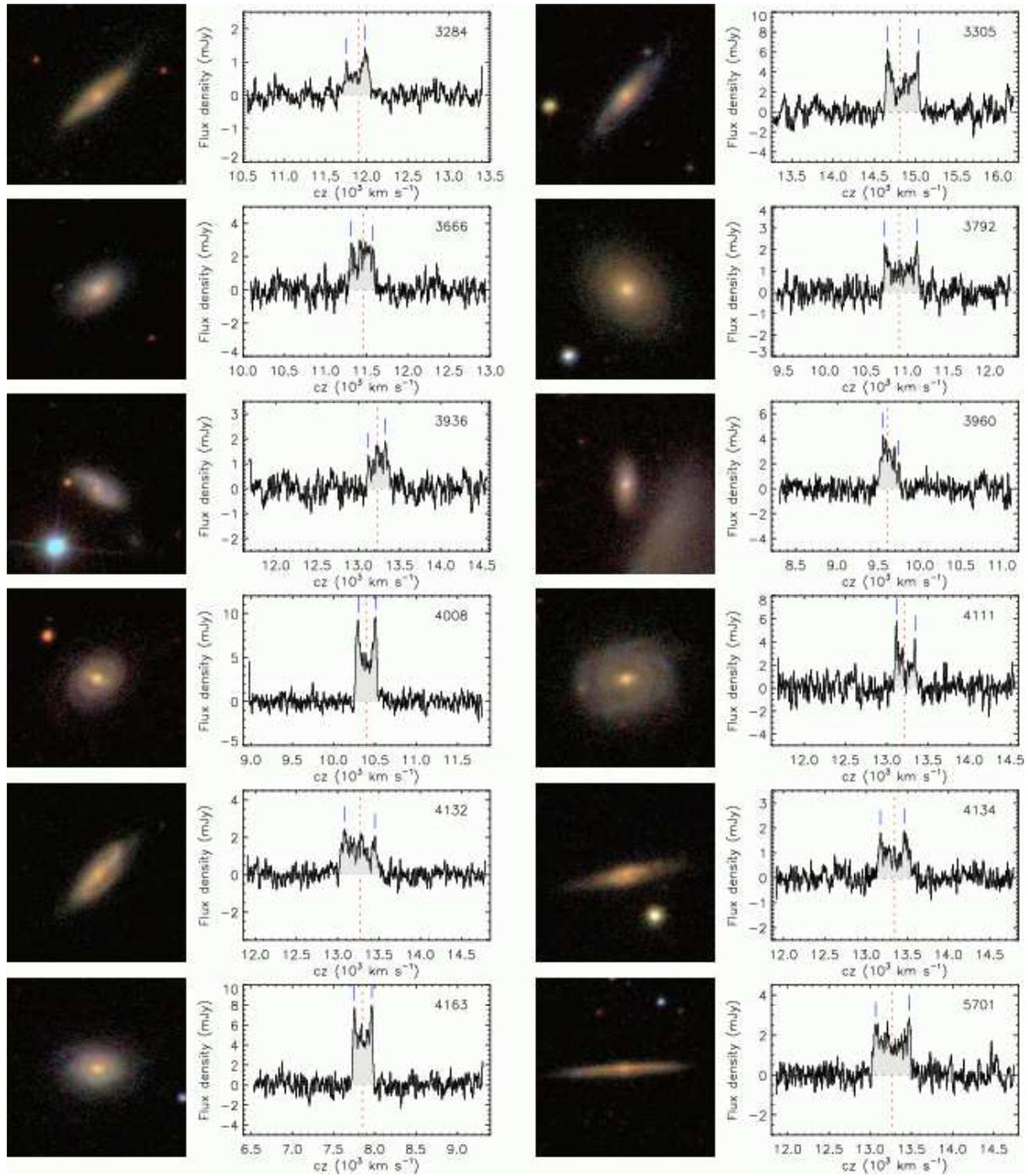}
\caption{SDSS postage stamp images (1 arcmin square) and
\hi-line profiles of the detections included in this final data
release, ordered by increasing GASS number (indicated in each spectrum). The \hi\ spectra are
calibrated, smoothed and baseline-subtracted. A dotted line and two
dashes indicate the heliocentric velocity corresponding to the SDSS
redshift and the two peaks used for width measurement,
respectively. This is a sample of the complete figure, which is
available in the online version of the article.}
\label{det}
\end{center}
\end{figure*}

\begin{figure*}
\includegraphics[width=16cm]{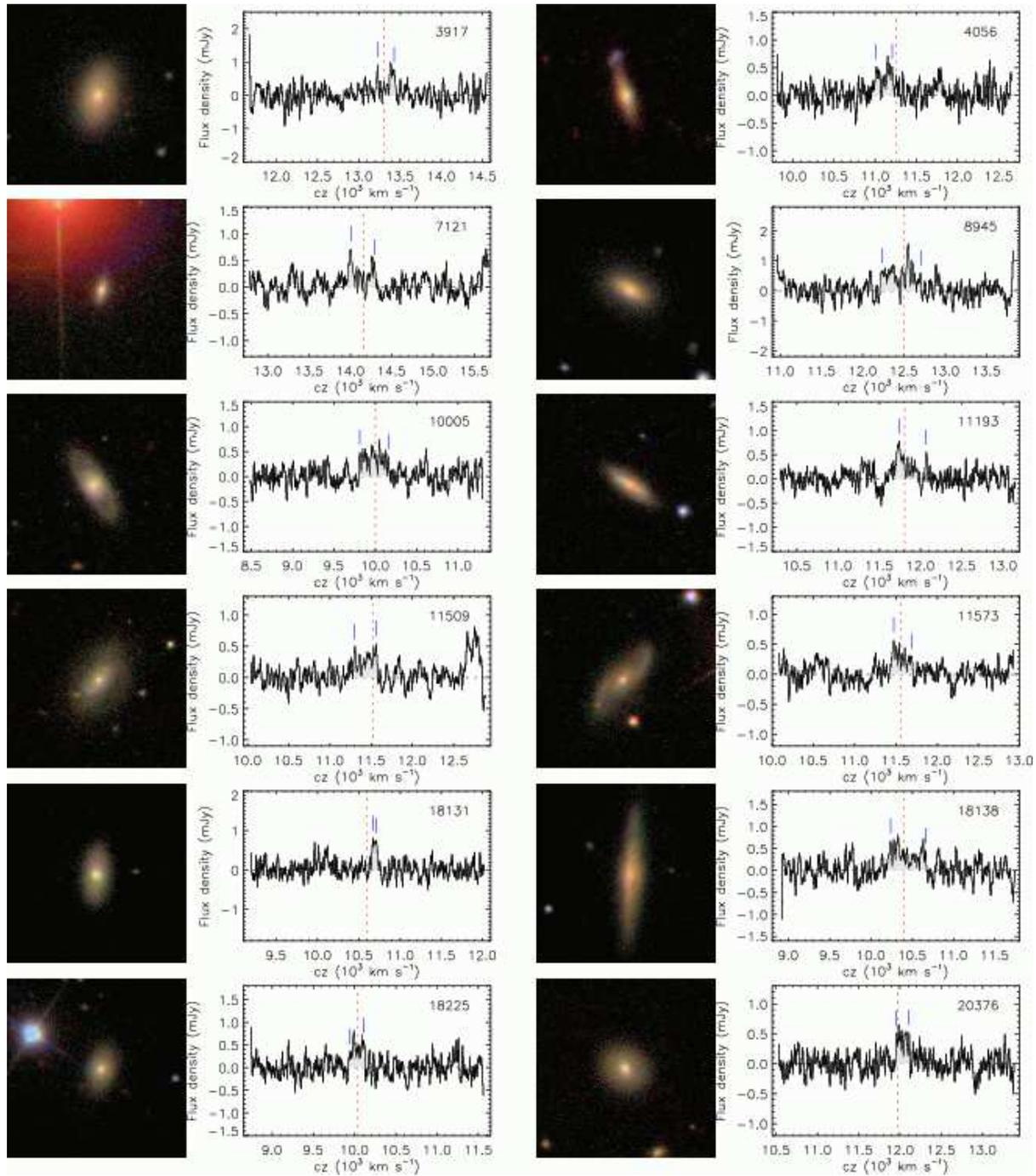}
\caption{Same as Figure~\ref{det} for marginal and/or confused
detections. Here galaxies are sorted by quality flag first (starting
with code 2 and increasing) and, within each category, by GASS number.
This is a sample of the complete figure, which is
available in the online version of the article.}
\label{marg_conf}
\end{figure*}

\begin{figure*}
\includegraphics[width=16cm]{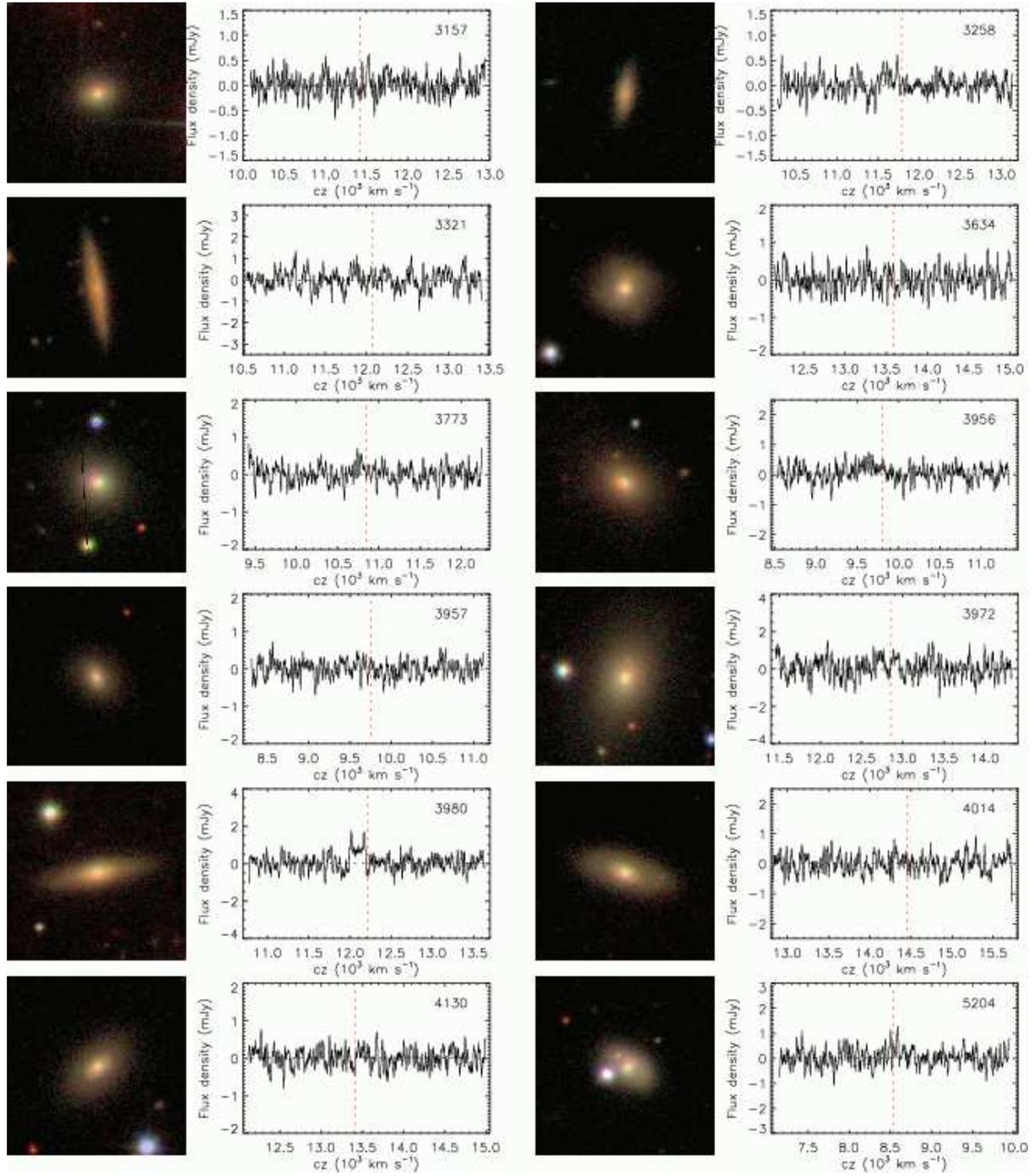}
\caption{Same as Figure~\ref{det} for non-detections. This is a sample of the complete figure, which is
available in the online version of the article.}
\label{ndet}
\end{figure*}

\noindent
{\bf SDSS postage stamps and \hi\ spectra.}\\
SDSS images and \hi\ spectra of the galaxies are presented here,
organized as follows: 
\hi\ detections with quality
flag 1 in Table~\ref{t_det} (Figure~\ref{det}), marginal and/or
confused detections with quality flag 2-5 (Figure~\ref{marg_conf}) 
and non-detections (Figure~\ref{ndet}).
The objects in each of these figures are ordered by 
increasing GASS number (indicated on the top right corner of each spectrum).
The SDSS images show a 1 arcmin square field, \ie\ only the central
part of the region sampled by the Arecibo beam (the half
power full width of the beam is \about 3.5\arcmin\ at the
frequencies of our observations). Therefore, companions that might be
detected in our spectra typically are not visible in the
postage stamps, but they are noted in Appendix~B.
The \hi\ spectra are always displayed over a 3000 \kms\ velocity
interval, which includes the full 12.5 MHz bandwidth adopted for our
observations. The \hi-line profiles are calibrated, smoothed 
(to a velocity resolution between 5 and 21 \kms\ for
the detections, as listed in Table~\ref{t_det}, or to
\about 15 \kms\ for the non-detections), and
baseline-subtracted. A red, dotted line indicates the heliocentric
velocity corresponding to the optical redshift from SDSS. 
In Figures~\ref{det}-\ref{marg_conf}, the shaded area and two vertical
dashes show the part of the profile that was integrated to
measure the \hi\ flux and the peaks used for width measurement,
respectively.

\section*{Appendix B: Notes on Individual Objects}\label{s_notes}

We list here notes for galaxies marked with an asterisk in
the last column of Tables~\ref{t_det} and \ref{t_ndet}.
The galaxies are ordered by increasing GASS number. In what follows, 
AA2 is the abbreviation for ALFALFA detection code 2.\\


\noindent
{\bf Detections (Table \ref{t_det})}\\

\noindent
{\bf 3666}  -- small blue companion \about 1 arcmin SW, SDSS J011759.89+153148.0 ($z=0.038248$), some contamination
 	       certain. There is also a blue galaxy \about 3 arcmin N, no SDSS redshift. \\
{\bf 3851}  -- offset from SDSS redshift, confused. This is a galaxy group, including a blue disk 
 	       \about 2 arcmin W (SDSS J013844.52+150331.1, $z=0.028769$, 8625 \kms) and two early-type galaxies 
 	       \about 1.5 arcmin SW (SDSS J013848.58+150141.2, $z=0.028044$) and \about 2 arcmin SE (SDSS J013854.76+150117.7,
 	       $z=0.027916$). \\
{\bf 3917}  -- marginal detection. Notice companion spiral galaxy \about 4 arcmin NW, SDSS J015742.52+132318.8 
 	       ($z=0.044431$). \\ 
{\bf 3936}  -- small blue disk \about 2 arcmin N has no SDSS redshift. \\
{\bf 3960}  -- spectacular pair of interacting galaxies in the foreground ($z=0.012$). AA2. \\
{\bf 3966}  -- blend/confused with blue companion \about 2 arcmin W, SDSS J020447.70+140147.8 ($z=0.030942$). \\
{\bf 3987}  -- detected blue companion \about 2 arcmin W, SDSS J021327.81+132806.1 ($z=0.041604$, 12473 \kms), confusion certain. \\
{\bf 4056}  -- high frequency edge uncertain, systematic error. Small blue cloud at the N edge of the galaxy, perhaps 
	       responsible for the peak at 11750 \kms? \\
{\bf 4111}  -- AA2. \\
{\bf 4136}  -- low frequency edge uncertain, systematic error. Blue companion 
 	       \about 40 arcsec SE, SDSS J015706.42+130926.9 ($z=0.032673$, 9795 \kms), confused. The blue galaxy
 	       \about 40 arcsec NE, SDSS J015706.42+131039.4, has $z=0.044781$. \\
{\bf 4165}  -- confused: blue companions \about 1.5 arcmin W (SDSS J015040.40+134106.1, $z=0.044814$, 13435 \kms)
 	       and 3 arcmin S (SDSS J015047.04+133824.1, $z=0.04437$, 13302 \kms); two other blue galaxies
 	       within 3 arcmin NE are in the background (SDSS J015057.74+134249.2, $z=0.050$ and 
 	       SDSS J015052.62+134318.6, $z=0.057$). \\
{\bf 5701}  -- two blue galaxies within 2.5 arcmin are in the background ($z=0.08$); small early-type galaxy 
 	       40 arcsec N has no SDSS redshift, but is unlikely to contaminate the signal. \\
{\bf 6015}  -- RFI spike at 1375 MHz (\about 9900 \kms). \\
{\bf 6679}  -- RFI feature at 1360 MHz (13300 \kms). Three blue galaxies \about 3 arcmin W: the large edge-on disk 
 	       (SDSS J130158.47+030602.6, $z=0.023386$ from NED) and its small companion to the N are in the
 	       foreground, and SDSS J130200.53+030550.1 is in the background ($z=0.079602$). \\
{\bf 7121}  -- near bright star. \\  
{\bf 7405}  -- asymmetric profile, uncertain width; several small galaxies within 2 arcmin, the only two with 
 	       redshifts are in the background ($z=0.056$ and $z=0.13$). \\
{\bf 7813}  -- blue companion \about 1.5 arcmin E, SDSS J151249.84+012827.7 ($z=0.0304$, 9114 \kms), is separated enough
 	       in velocity not to cause any confusion (there is a small peak at the right velocity, but it
 	       is present in one polarization only). Also, blue companion 3.6 arcmin NW, SDSS J151233.08+013017.3 
 	       ($z=0.029178$). \\
{\bf 8096}  -- low frequency edge uncertain, systematic error. Small blue companion \about 1.5 arcmin SW, 
 	       SDSS J085249.56+030823.9 ($z=0.034776$), some contamination certain. \\
{\bf 8634}  -- possibly confused with blue galaxy \about 2 arcmin N, SDSS J101322.37+050312.7, 
 	       no optical redshift (photometric redshift $z=0.042$). \\
{\bf 8945}  -- blue companion \about 3 arcmin W, SDSS J105303.39+042036.5 ($z=0.041924$, 12568 \kms), some contamination
 	       likely; the blue galaxy \about 1 arcmin NW has $z=0.066$. \\ 
{\bf 9615}  -- RFI spike at 1375 MHz (\about 9900 \kms), 2 channels replaced by interpolation. No companions within 3 arcmin,
 	       galaxy \about 2.5 arcmin SW is in the background (SDSS J142955.29+032157.1, $z=0.168$). \\
{\bf 9942}  -- stronger in polarization A. Blend/confused with edge-on disk 0.4 arcmin NE, SDSS J144326.39+042308.2, 
	       $cz=7879$ \kms from NED. Blue galaxy \about 2.5 arcmin NE, SDSS J144332.81+042423.1, has $z=0.071$. \\
{\bf 10005} -- blue disk \about 3 arcmin SW, SDSS J145259.66+033013.2, is in the background ($z=0.045$). \\
{\bf 10032} -- three galaxies \about 1 arcmin N, 2 arcmin E and 2.5 arcmin SE are in the background ($z=0.094$, 
               0.209 and 0.094, respectively). \\
{\bf 11086} -- 2.7 Jy continuum source at 5 arcmin, standing waves. \\
{\bf 11092} -- galaxy pair, the companion is a blue spiral \about 15 arcsec SE, SDSS J231340.49+140115.5  
	       ($z=0.040436$, 12122 \kms). Notice another two disk galaxies at the same redshift, \about 2.5 arcmin SW
	       (SDSS J231334.71+135912.4, $z=0.039767$) and \about 3 arcmin NW (SDSS J231330.39+140349.7, $z=0.039527$). \\
{\bf 11193} -- uncertain profile; early-type companion \about 1.5 arcmin E, SDSS J231328.01+141611.3 ($z=0.038994$, 11690 \kms); 
	       another companion \about 4 arcmin NE, SDSS J231331.44+141938.7 ($z=0.039231$, 11761 \kms), significant 
	       contamination unlikely. Small galaxy 40 arcsec W has $z=0.150$. Better in polarization B. \\
{\bf 11291} -- companion of GASS 11292, \about 2.5 arcmin SW; strong contamination is unlikely, see note for GASS 11292. \\
{\bf 11292} -- most of the signal comes from GASS 11291 \about 2.5 arcmin NE, as can be seen by comparing the two profiles. \\
{\bf 11312} -- galaxy triplet, \hi\ signal is most likely a blend. The two companions are disk galaxies 1.9 arcmin NE
	       (SDSS J231229.22+135632.1, $z=0.034137$) and 2.3 arcmin N (SDSS J231224.51+135704.5, $z=0.034135$).\\
{\bf 11347} -- most likely confused/blend with large spiral \about 2 arcmin W, SDSS J231639.26+153516.2 ($z=0.038807$ from NED). \\
{\bf 11434} -- small companion \about 2.5 arcmin S, SDSS J232328.01+140530.2 ($z=0.041497$), some contamination possible. \\
{\bf 11435} -- small companions \about 2 arcmin NW (SDSS J232314.91+141817.8, $z=0.043379$) and \about 2.5 arcmin S
	       (SDSS J232318.65+141446.6, $z=0.044175$), some contamination possible. Large spiral galaxy 
	       \about 3 arcmin NE is in the foreground ($z=0.026$). \\
{\bf 11509} -- high frequency edge uncertain, systematic error. Detected (part of) blue companion \about 1.7 arcmin NW,
 	       SDSS J232403.09+145137.7 ($z=0.042698$, 12801 \kms). \\
{\bf 11573} -- stronger in polarization B. Early-type companion 2 arcmin E, SDSS J233019.67+132657.3 ($z=0.039838$);
 	       the early-type galaxy \about 1 arcmin N, SDSS J233013.51+132801.6, has $z=0.041588$ (12468 \kms). \\
{\bf 11669} -- edge-on galaxy \about 1 arcmin SE, SDSS J232715.24+152752.4 ($z=0.046110$ from NED), some contamination 
 	       possible (although the profile is consistent with the fact that the target is almost face-on). AA2. \\
{\bf 12062} -- reddish companion \about 2 arcmin NE, same redshift ($z=0.036556$), and two small galaxies \about 30 arcsec S, no 
 	       redshifts; small contamination possible. \\
{\bf 13159} -- no obvious companion within the beam, however notice two small, blue smudges \about 1 and 1.5 arcmin E,
 	       without optical redshifts. \\
{\bf 13618} -- blend with companion 1 arcmin S, SDSS J135621.74+043606.0 ($z=0.03382$, 10139 \kms).\\
{\bf 13674} -- AA2. \\
{\bf 14247} -- small companion \about 2 arcmin SW, SDSS J080523.82+355454.5 ($z=0.033211$), some contamination possible. \\
{\bf 15257} -- uncertain profile; most likely confused/blend with blue galaxy \about 45 arcsec W, SDSS J104802.72+060103.7, 
 	       without optical redshift. \\
{\bf 17622} -- disk galaxy \about 1.5 arcmin SW is in the background ($z=0.061$). AA2. \\
{\bf 17673} -- confused/blend with blue companion \about 3 arcmin E, SDSS J110009.92+102214.1 ($z=0.036759$, 11020 \kms);
 	       small galaxy \about 1 arcmin SE has $z=0.092$. \\
{\bf 18084} -- detected blue companion \about 2 arcmin W, SDSS J115104.26+085225.0 ($z=0.036558$, 10960 \kms). \\
{\bf 18131} -- two blue galaxies in the background, one \about 1 arcmin N (SDSS J120446.89+092617.6, $z=0.069$) and 
 	       one 3 arcmin S (SDSS J120446.35+092222.2, $z=0.041$). Notice however blue, low surface brightness (LSB) 
	       galaxy 1 arcmin S, SDSS J120445.64+092426.7, without optical redshift. Confused? \\
{\bf 18138} -- early-type companion \about 3 arcmin W, SDSS J120227.14+085548.2 ($z=0.034643$),
 	       significant contamination unlikely. Stronger in polarization B. \\
{\bf 18225} -- blue disk \about 1 arcmin W, SDSS J120507.73+103352.6, without optical redshift. Small blue galaxy
 	       \about 2 arcmin E, SDSS J120517.65+103320.5, has $z=0.023$. Notice large early-type companion \about 3.5 arcmin N,
 	       SDSS J120514.04+103647.6 ($z=0.033449$). Three other galaxies
 	       \about 3 arcmin away in the W quadrant are in the background ($z=0.09$). \\
{\bf 19672} -- galaxy pair. \\
{\bf 19989} -- several small galaxies around without optical redshifts; galaxy \about 2.5 arcmin SW, 
 	       SDSS J085419.08+081057.2, has $z=0.096$. \\
{\bf 20041} -- large, blue companion 3.3 arcmin NE, SDSS J091437.31+080702.0 ($z=0.031015$ from NED). The companion
 	       is detected by ALFALFA (AGC 191126) with \whi $=402$ \kms and flux of 3.09 Jy \kms. Confused? \\
{\bf 20376} -- polarization mismatch (clear, overlapping signal in both polarizations, but offset by 1 MHz).
 	       The signal is most likely confused/blend with that of a blue spiral \about 2.5 arcmin NW, 
	       SDSS J095407.95+103625.6 (also AGC 193987, detected by ALFALFA; $z=0.040392$, 12109 \kms). Notice also GASS 20445
 	       \about 3 arcmin E ($z=0.039708$, 11904 \kms; non-detection in this release). \\
{\bf 23070} -- spiral galaxy 3 arcmin W has $z=0.109$. \\
{\bf 23496} -- RFI spikes near 1352.5 MHz (\about 15,000 \kms). Small companion \about 1 arcmin E, SDSS J105725.50+120638.9 
	       ($z=0.047348$), and galaxy \about 30 arcsec NW without SDSS redshift; some contamination possible. AA2. \\
{\bf 23703} -- small blue galaxy \about 2 arcmin S, no optical redshift. \\
{\bf 23739} -- blue companion \about 3 arcmin SW, SDSS J113655.36+115053.9 ($z=0.034412$, 10316 \kms), separated enough
 	       in velocity from the target.  \\
{\bf 23781} -- confused/blend with large blue spiral \about 2 arcmin NW, SDSS J114206.64+113216.0 ($z=0.042924$, 12868 \kms).\\
{\bf 23789} -- most likely confused/blend with blue companion \about 2.7 arcmin E, SDSS J114154.89+123030.7 ($z=0.034531$, 10352 \kms). \\
{\bf 23815} -- small galaxy \about 20 arcsec SW has no redshift; blue galaxy \about 2 arcmin E is in the background ($z=0.052$). \\
{\bf 24496} -- small blue companion \about 2 arcmin N, SDSS J111809.86+074845.7 ($z=0.041832$), some contamination certain. \\
{\bf 25215} -- several galaxies to the S, all in the background. \\
{\bf 25721} -- small blue galaxy \about 1.7 arcmin S, SDSS J155507.68+092848.6, no optical redshift. \\
{\bf 26406} -- small galaxy \about 2.5 arcmin W has $z=0.044$; smudge \about 1 arcmin E has no SDSS redshift. \\
{\bf 26407} -- RFI spike at 1352.6 MHz (15,000 \kms). Edge-on galaxy \about 2 arcmin NE is in the background ($z=0.086$). \\
{\bf 26586} -- notice two blue, edge-on disks \about 4 arcmin from the target and with similar redshifts:
 	       SDSS J103624.87+130827.0 (4 arcmin SE, $z=0.034084$) and SDSS J103619.24+131317.5 
 	       (3.5 arcmin NE, $z=0.033366$). \\
{\bf 28062} -- most likely blend with small companion \about 1.5 arcmin E, SDSS J122807.37+081057.3 
 	       ($z=0.037407$, 11214 \kms), which is exactly centered on the highest peak. Also notice companion
 	       galaxy \about 3.5 arcmin NE. \\
{\bf 28317} -- companion \about 2 arcmin NW, SDSS J154403.74+274152.5 ($z=0.031411$), but there is no hint of detection on the side 
	       away from GASS~28317, so contamination is unlikely. Notice however disk galaxy next to it, without optical redshift.\\ 
{\bf 28703} -- AA2. \\
{\bf 31095} -- AA2. \\
{\bf 32308} -- AA2. \\
{\bf 33214} -- high frequency edge uncertain, systematic error. The disk galaxy \about 2.5 arcmin S has $z=0.050$. \\
{\bf 33737} -- disturbed, no companions within the beam, large offset from SDSS redshift ($z=0.026869$, 8055 \kms).\\
{\bf 38198} -- several galaxies within 3 arcmin in the background ($z=0.097$). \\
{\bf 38458} -- uncertain profile; blend: connected to large companion \about 40 arcsec E, SDSS J140606.72+123013.6 
 	       (also GASS 25575, not detected in DR1; $z=0.037966$, 11382 \kms);
 	       notice also small companion \about 1.5 arcmin W, SDSS J140557.71+123016.6 ($z=0.039257$, 11769 \kms). \\
{\bf 39082} -- blue LSB galaxy \about 1 arcmin SE, no optical redshift (photo-$z=0.037$), possible contamination. AA2.  \\
{\bf 40495} -- low frequency edge uncertain, systematic error; stronger in polarization A. \\
{\bf 40502} -- 163 mJy continuum source at 1 arcmin, standing waves. \\
{\bf 41718} -- detected blue companion in board 3, \about 1365 MHz (\about 12150 \kms), most likely the very blue 
	       galaxy \about 1 arcmin NW, SDSS J144334.78+083432.3 (no optical redshift); galaxy \about 2.5 arcmin W, 
 	       SDSS J144328.85+083248.9, has $z=0.033037$ (9904 \kms). \\
{\bf 41863} -- interacting pair of blue galaxies: the companion is \about 40 arcsec E, SDSS J151031.62+072500.2 
 	       (c$z=9597$ \kms from NED).\\
{\bf 41869} -- detected blue companion, SDSS J150921.31+070631.4, \about 2 arcmin N ($z=0.037367$, 1369.24 MHz, 11200 \kms);
 	       galaxy \about 1 arcmin NE is in the background ($z=0.078$). \\
{\bf 42191} -- profile edges uncertain, systematic error. \\
{\bf 42233} -- several galaxies within 3 arcmin in the background ($z>0.08$). \\
{\bf 44856} -- tiny blue galaxy \about 1.5 arcmin S, SDSS J135409.08+243200.3, unknown redshift. \\
{\bf 48356} -- high frequency edge uncertain, systematic error. Interacting with SDSS J111113.00+284242.7, \about 1 arcmin N 
	       ($z=0.029366$, 8804 \kms); several other galaxies with similar redshift within 3 arcmin. \\
{\bf 48518} -- low frequency edge uncertain, systematic error; uncertain profile.
 	       Blend with large spiral 2 arcmin S, SDSS J111750.61+263732.8 ($z=0.027048$, 8109 \kms); there is
 	       also a small companion \about 1 arcmin N, SDSS J111751.46+264035.3 ($z=0.026349$, 7899 \kms). \\
{\bf 48521} -- small blue galaxy \about 30 arcsec E, SDSS J111740.84+263502.1, unknown redshift, possible confusion. AA2.\\
{\bf 48604} -- small blue companion \about 1.5' N, SDSS J112746.74+265909.7 ($z=0.033782$, 10129 \kms), some contamination
 	       certain. Several smaller galaxies within \about 2 arcmin, either in the background or without 
 	       SDSS redshift. AA2. \\
{\bf 48994} -- two small blue companions \about 2.5 arcmin NE, SDSS J114225.77+301549.5 and
 	       SDSS J114227.14+301552.6 (both have $z=0.033$), likely adding very little to the 
 	       signal (given their size and distance to the target). \\
{\bf 49386} -- small spiral \about 2 arcmin SW is in the background ($z=0.080$). \\
{\bf 49727} -- low frequency edge uncertain, systematic error. Galaxy pair, separation 4 arcsec (from NED); \hi\ signal 
	       also blended with that of UGC 7064 (\about 1 arcmin S, $z=0.024916$, 1385.88 MHz, face-on blue galaxy,
 	       which is responsible for the low velocity peak) and likely with that of 
 	       SDSS J120445.26+310927.8 (blue galaxy 2 arcmin S, $z=0.026637$, 1383.55 MHz). \\
{\bf 50404} -- small companion \about 2 arcmin SW, SDSS J123400.02+280641.8 ($z=0.040307$), some contamination
 	       possible. Spiral \about 3 arcmin NE is in the background ($z=0.084$). \\
{\bf 50406} -- low frequency edge uncertain, systematic error. \\
{\bf 51150} -- small companion \about 1 arcmin S, SDSS J132301.23+270558.8 ($z=0.034507$); notice also blue companion 
 	       3.7 arcmin SE, SDSS J132309.49+270359.2 ($z=0.034215$); significant contamination unlikely. \\
{\bf 51161} -- AA2. \\
{\bf 51334} -- small companion \about 1.5 arcmin N, SDSS J075331.69+140237.3 ($z=0.029093$), some 
 	       contamination possible. AA2. \\
{\bf 51580} -- red companion \about 2 arcmin SW, SDSS J080359.21+150343.1 ($z=0.039$), significant contamination 
 	       unlikely. Several small galaxies nearby without SDSS redshift. \\
{\bf 51899} -- blend with two companions, a blue edge-on disk 2 arcmin S (SDSS J083131.00+192042.6, $z=0.039271$)
 	       and a large galaxy \about 2.5 arcmin E (SDSS J083140.72+192307.8, $z=0.038759$); there is also a small
 	       bluish galaxy \about 15 arcsec NE without optical redshift. \\
{\bf 52045} -- disturbed (tidal tail or companion to the S). \\
{\bf 52297} -- companion \about 2 arcmin NE, SDSS J085724.03+204237.8 ($z=0.032874$), plus several galaxies nearby without
 	       SDSS redshifts; some contamination likely. \\
{\bf 55541} -- small blue galaxy \about 1 arcmin SW has $z=0.048$, no comtamination problems. \\
{\bf 55745} -- stronger in polarization B. \\
{\bf 56509} -- 111 mJy continuum source at 2 arcmin, standing waves. Small blue companion \about 2 arcmin N, 
 	       SDSS J085047.04+115102.8 ($z=0.029322$), some contamination likely. \\
{\bf 56662} -- a few small galaxies within 3 arcmin, all in the background. \\


\noindent
{\bf Non-detections (Table \ref{t_ndet})}\\

\noindent
{\bf 3157}  -- small face-on, spiral companion \about 2 arcmin E, SDSS J003042.29+145610.4 ($z=0.038491$, 11539 \kms). \\
{\bf 3258}  -- perhaps hint of galaxy signal; blue galaxy \about 2.5 arcmin SW has $z=0.076$. \\
{\bf 3972}  -- edge-on companion \about 3 arcmin SW, SDSS J020530.66+143652.7 ($z=0.042305$, 12683 \kms);
 	       hint of signal centered at 12800 \kms\ is in polarization B only. \\
{\bf 3980}  -- detected LSB companion \about 1.5 arcmin S, SDSS J021424.66+121836.7 ($z=0.040399$, 12111 \kms, 
 	       in much better agreement with SDSS redshift). Stronger in polarization A. \\       
{\bf 4014}  -- small companion \about 3.5 arcmin NE, SDSS J020732.08+130338.3  ($z=0.048163$, 14439 \kms). \\
{\bf 4130}  -- blue companion \about 3.5 arcmin W, SDSS J015706.42+131039.4 ($z=0.044781$); several other galaxies
 	       within 3 arcmin, with redshifts significantly different from GASS~4130 or unknown. \\
{\bf 5204}  -- blue companion \about 3 arcmin SE, SDSS J102800.60+023414.1 ($z=0.028467$, 8534 \kms) also not detected. \\
{\bf 7310}  -- the two disk galaxies 3 arcmin S and \about 3.5 arcmin NE have $z>0.05$. \\
{\bf 8953}  -- blue companion \about 3 arcmin NE, SDSS J105251.60+041109.3 ($z=0.043311$, 12984 \kms), marginally detected? \\
{\bf 8971}  -- two companions: edge-on disk \about 1.5 arcmin NW, SDSS J104832.28+044838.1 ($z=0.033723$, 10110 \kms)
 	       and face-on, blue spiral \about 3 arcmin NW, SDSS J104827.34+044931.7 ($z=0.034128$, 10231 \kms), also
 	       not detected. The small galaxy \about 3 arcmin SE, SDSS J104847.35+044605.5, has $z=0.026$. \\
{\bf 9702}  -- small galaxy \about 1 arcmin W, SDSS J144039.22+032250.3 ($z=0.030114$, 9028 \kms); several other 
 	       galaxies within \about 3 arcmin with redshifts $z<0.028$ or $z>0.089$. \\
{\bf 10211} -- blue galaxy \about 1.5 arcmin S has $z=0.093$. \\
{\bf 11080} -- double nucleus; detected blue companion \about 2 arcmin N 
 	       in board 3, SDSS J225609.41+130551.4 ($z=0.037436$, 1369.15 MHz). \\
{\bf 11249} -- companion of GASS 11257, \about 2 arcmin E (SDSS J230806.95+152520.2, $z=0.036716$, see next note);
 	       large early-type galaxy \about 0.5 arcmin N without optical redshift. \\
{\bf 11257} -- companion of GASS 11249, \about 2 arcmin W (SDSS J230757.92+152455.2, $z=0.03623$, see previous note); 
 	       large early-type galaxy \about 2.5 arcmin W without optical redshift. \\
{\bf 11284} -- perhaps hint of galaxy signal. \\
{\bf 11395} -- small companion \about 2 arcmin S, SDSS J232336.22+133706.0 ($z=0.042373$, 12703 \kms). \\
{\bf 11410} -- detected companion without redshift? perhaps the small LSB galaxy \about 0.5 arcmin NW, 
 	       SDSS J232220.71+135957.4; the signal is significantly stronger in polarization A,
 	       although no RFI is visible at that frequency. \\
{\bf 11544} -- AA2. Marginally detected disk galaxy \about 2 arcmin NW, SDSS J232531.72+152211.6 ($z=0.040311$, 12085 \kms). \\
{\bf 11567} -- two companions, a large spiral 2 arcmin W, SDSS J233011.60+132656.2 ($z=0.038729$, 11611 \kms)
 	       and a small one \about 2 arcmin SE, SDSS J233024.64+132531.6 ($z=0.039084$, 11717 \kms); notice also
 	       the early-type galaxy \about 2 arcmin NW, SDSS J233013.51+132801.6 ($z=0.041588$, 12468 \kms). \\
{\bf 11568} -- large early-type galaxy \about 2 arcmin SE, SDSS J233019.67+132657.3 ($z=0.039838$) and a spiral
 	       \about 1.5 arcmin S, SDSS J233011.60+132656.2 ($z=0.038729$). \\
{\bf 11585} -- marginally detected blue companion \about 3 arcmin N, SDSS J232519.80+142419.5 ($z=0.042036$, 12602 \kms). \\
{\bf 11636} -- two blue companions: SDSS J232336.90+151532.2 \about 2 arcmin NE ($z=0.043298$,
 	       12980 \kms) and GASS 11494 \about 2.5 arcmin S (SDSS J232335.34+151148.7, $z=0.042709$, 12804 \kms,
 	       DR2 detection). The \hi\ signal is most likely a blend of the two. \\
{\bf 11791} -- marginally detected blue companion \about 1.5 arcmin SE, SDSS J235205.28+144403.6 ($z=0.0450$, 13491 \kms);
 	       another two galaxies with similar redshifts \about 2 arcmin N (SDSS J235158.07+144711.1, $z=0.046598$, 
 	       13970 \kms) and \about 3.5 arcmin SW (SDSS J235148.64+144241.3, $z=0.046466$, 13930 \kms). \\
{\bf 11892} -- three galaxies within 3 arcmin with $z>0.09$, and two edge-on disks \about 3 arcmin S without optical redshifts. \\
{\bf 11903} -- two small galaxies within 0.5 arcmin without optical redshift. \\
{\bf 12452} -- companion \about 2 arcmin S, SDSS J112003.92+040830.2 ($z=0.049371$, 14801 \kms); three other galaxies
 	       within 2.5 arcmin with $z=0.15$. \\
{\bf 12967} -- companion of GASS 12970, \about 2 arcmin S (SDSS J123553.79+054539.8, $z=0.041788$, DR2 non-detection).
 	       Small galaxy \about 2.5 arcmin S, SDSS J123556.38+054459.2 ($z=0.041189$, 12348 \kms) also not detected.
 	       \tmax\ not reached; perhaps hint of galaxy signal, but stronger in polarization A. \\
{\bf 14017} -- small blue cloud to the NW edge of the galaxy; galaxy pair \about 4 arcmin SW in the foreground ($z=0.016$). \\
{\bf 14260} -- small blue galaxy \about 2 arcmin S has $z=0.145$. \\
{\bf 16756} -- edge-on disk \about 40 arcsec SE and small blue galaxy \about 1 arcmin SW, both without optical redshift. \\
{\bf 18004} -- barred spiral galaxy \about 40 arcsec SE and small galaxy \about 40 arcsec W, both without optical redshifts;
 	       small early-type companion \about 2.5 arcmin E, SDSS J115145.97+084531.2 ($z=0.035926$) also not detected. \\
{\bf 18220} -- perhaps hint of galaxy signal. Early-type companion \about 2.5 arcmin W, SDSS J120527.47+104204.4 
 	       ($z=0.035443$, 10626 \kms); the blue galaxy \about 2.5 arcmin NW, SDSS J120530.37+104313.8, has $z=0.063$. \\
{\bf 19274} -- AA2. Small companion \about 3 arcmin N, SDSS J081622.22+260241.1 ($z=0.045113$) also not detected. \\
{\bf 20149} -- hint of galaxy signal. \\
{\bf 20165} -- detected companion (GASS 20133, DR1 detection) \about 1 arcmin E (SDSS J093236.58+095025.9, $z=0.048884$, 14655 \kms). \\
{\bf 20445} -- companion of GASS 20376, \about 3 arcmin W (SDSS J095416.83+103457.5, $z=0.039938$, see detections in this release). \\
{\bf 23029} -- three companions: SDSS J102714.26+110340.0, \about 2 arcmin W ($z=0.032969$), SDSS J102710.59+110116.2,
 	       \about 2.5 arcmin S ($z=0.032657$) and SDSS J102707.78+110038.5, \about 3 arcmin S ($z=0.032367$). \\
{\bf 23102} -- perhaps hint of galaxy signal. \\
{\bf 23203} -- detected companion, probably the LSB galaxy \about 3 arcmin E, SDSS J103602.58+121118.3
 	       ($z=0.038124$, 11429 \kms), but notice also blue smudge \about 1.5 arcmin NE without optical redshift. \\
{\bf 23457} -- early-type galaxy \about 1 arcmin NE without optical redshift. \\
{\bf 23531} -- RFI at 1370 MHz (\about 11000 \kms) visible in final spectrum.  \\
{\bf 25057} -- companion of GASS 25115 (SDSS J152112.78+303928.5, \about 1.8 arcmin SE, see next note), also not detected; several
 	       other galaxies within 3 arcmin in the background ($z>0.07$). \\
{\bf 25115} -- companion of GASS 25057 (SDSS J152106.26+304036.9, see previous note), also not detected; several
 	       other galaxies within 3 arcmin in the background ($z>0.07$). \\
{\bf 25213} -- detected companion, most likely SDSS J131221.39+114022.8, 3 arcmin S ($z=0.030132$, 9033 \kms).
 	       Two other galaxies within the beam with sligthly higher redshift: SDSS J131229.71+114432.7, 
 	       \about 2 arcmin NE ($z=0.030916$, 9268 \kms) and GASS 25214 (SDSS J131232.81+114344.2, $z=0.031105$, 
	       9325 \kms), 2.5 arcmin E, which was not detected in DR1. \\
{\bf 25682} -- three galaxies within \about 3.5 arcmin at slightly higher redshift ($z=0.042$), plus others without optical redshifts. \\
{\bf 26017} -- large early-type companion \about 2.5 arcmin SW, SDSS J095634.77+110947.4 ($z=0.041272$, 12373 \kms),
 	       and small disk galaxy \about 2 arcmin W, SDSS J095633.88+111058.2 ($z=0.04049$, 12139 \kms). \\
{\bf 26503} -- large early-type companion 3 arcmin SE, SDSS J102323.75+125006.1 ($z=0.032486$, 9739 \kms); other
 	       galaxies within 3 arcmin are in the background or have no redshift. \\
{\bf 28030} -- perhaps hint of galaxy signal, but present in polarization B only. \\
{\bf 28327} -- next to bright star; two companion spirals \about 3.5 arcmin E (SDSS J154145.55+275917.8, $z=0.032026$)
 	       and 4 arcmin N (SDSS J154135.57+280258.6, $z=0.031891$, AA2). \\
{\bf 28348} -- companion \about 2 arcmin SE, SDSS J154100.75+281922.9 ($z=0.032234$, 9664 \kms); the small blue galaxy
 	       \about 2 arcmin E has $z=0.066$. \\
{\bf 30746} -- galaxy \about 1 arcmin NW has $z=0.079$. \\
{\bf 31478} -- hint of galaxy signal; two galaxies \about 2 arcmin W and \about 2 arcmin NW have $z=0.063$. \\
{\bf 33469} -- small blue galaxy \about 2.5 arcmin E is in the foreground ($z=0.005$). \\
{\bf 33777} -- detected blue companion \about 2 arcmin W, SDSS J100240.68+323749.4 ($z=0.045315$, 13585 \kms);
 	       several other galaxies within 3 arcmin with redshifts between 0.048 and 0.052. \\
{\bf 35475} -- face-on spiral galaxy \about 3 arcmin SE, SDSS J125941.30+283025.9 ($z=0.027566$, 8264 \kms), also not detected. \\
{\bf 35497} -- hint of galaxy signal. Four galaxies within 3.5 arcmin are in the background ($z>0.06$). \\
{\bf 39407} -- blue companion \about 2 arcmin SE, SDSS J152248.12+083148.0 ($z=0.036607$, 10975 \kms), and three
 	       galaxies \about 2.5 arcmin away with $z=0.034-0.035$ (SDSS J152236.57+083447.5, 
 	       SDSS J152249.21+083337.9, SDSS J152244.22+083013.3). \\
{\bf 41723} -- blue companion \about 3.5 arcmin NE, SDSS J144610.59+085807.7 ($z=0.029595$, 8872 \kms); also notice two
 	       blue disks \about 2.5 arcmin N without optical redshifts. \\
{\bf 44021} -- small blue galaxy \about 70 arcsec NE, SDSS J134235.17+301547.2, 
 	       has $z=0.124$; very blue galaxy \about 2 arcmin SW, SDSS J134226.53+301311.0, has no optical redshift. \\
{\bf 45940} -- small bluish cloud to the N of the galaxy. Large red companion \about 2.5 arcmin NE, SDSS J142758.18+263016.2 
	       ($z=0.032298$) and two galaxies \about 3 arcmin SE, SDSS J142759.07+262754.1 
 	       ($z=0.031056$) and SDSS J142759.98+262805.9 (no optical redshift). \\
{\bf 48160} -- feature at \about 13250 \kms is present in both polarizations; detected
 	       perhaps the small blue galaxy \about 1 arcmin S (SDSS J111203.29+274951.2) without optical redshift?
{\bf 48205} -- disk galaxy 2 arcmin N is in the foreground ($z=0.037$); small galaxies within 1.5 arcmin without optical redshifts. \\
{\bf 48544} -- RFI spike at 1350 MHz (\about 15600 \kms) visible in final spectrum. Small blue companion \about 3 arcmin E, 
	       SDSS J112053.20+271816.7 ($z=0.047646$, 14284 \kms). \\
{\bf 50866} -- small disk galaxy \about 2 arcmin SE, SDSS J125614.26+274856.0 ($z=0.022243$, 6668 \kms); a few other
 	       small galaxies within 3 arcmin in the background ($z>0.08$). \\
{\bf 51462} -- small, blue companion \about 2 arcmin NW, SDSS J075555.63+141317.0 ($z=0.03625$, 10867 \kms) also not detected. \\
{\bf 53269} -- smaller galaxy almost superimposed is in the background (SDSS J093115.52+263255.8, $z=0.058$). \\
{\bf 54240} -- two edge-on galaxies nearby, SDSS J102254.55+243639.4 (\about 20 arcsec NE, $z=0.046341$, 13893 \kms)
 	       and SDSS J102248.59+243622.3 (\about 1 arcmin W, no optical redshift). \\
{\bf 54577} -- companion \about 2.5 arcmin W, SDSS J103007.19+273436.7 ($z=0.047206$). \\
{\bf 56320} -- detected companion, large spiral \about 1 arcmin N, SDSS J080343.91+100306.2 ($z=0.034116$, 10228 \kms);
	       also notice small blue galaxy \about 2.5 arcmin W, SDSS J080332.99+100259.1 ($z=0.034658$). \\
{\bf 56612} -- three large, disky companions: SDSS J090320.38+134142.0, \about 3 arcmin E ($z=0.029988$),
 	       SDSS J090313.10+134444.1, \about 3 arcmin NE ($z=0.028401$) and SDSS J090254.93+133938.4,
	       \about 4 arcmin SW ($z=0.029838$); the small galaxy \about 40 arcsec W is in the background ($z=0.102$). \\
{\bf 56650} -- perhaps hint of galaxy signal (not well centered on SDSS redshift).\\


\onecolumn
\begin{landscape}
\begin{table*}
\small
\centering
\caption{SDSS and UV Parameters.}
\label{t_sdss}
\begin{tabular}{rclccccccccccccc}
\hline\hline
   &  &  &  & Log \Mst & \Rinz & $R_{50}$ & $R_{90}$ & Log \must  & ext$_r$ & r & (b/a)$_r$ & incl &\nuvr & T$_{NUV}$ & \tmax \\
GASS  & SDSS ID & Other name & $z_{\rm SDSS}$ & (\Msun) & (\arcsec) & (\arcsec)&(\arcsec) & (\Msun~kpc$^{-2}$)& (mag)& (mag) &   & (deg) & (mag)& (sec) & (min) \\
(1)  & (2)  & (3)  & (4)  & (5) & (6) & (7) & (8)  & (9)  &  (10) & (11) & (12) & (13) & (14) & (15) & (16)\\
\hline
 11892 & J000200.82+150132.1 &  ...	     &0.0357 & 10.54 &  3.12 &  3.37 &  9.96 &  8.99 &  0.13 & 15.03 &   0.820 &   36 &  5.60 &    1705 &   49 \\
 11903 & J000458.72+154018.2 &  ...	     &0.0373 & 10.17 &  2.16 &  2.11 &  5.99 &  8.90 &  0.19 & 15.93 &   0.411 &   69 &  5.49 &     260 &   71 \\
 11910 & J000632.57+154004.7 &  ...	     &0.0370 & 10.28 &  2.97 &  3.10 &  8.53 &  8.74 &  0.18 & 15.67 &   0.555 &   58 &  3.69 &     260 &   68 \\
 12030 & J001842.68+151142.6 &  ...	     &0.0372 & 10.65 &  2.41 &  2.44 &  7.63 &  9.29 &  0.13 & 14.85 &   0.724 &   45 &  5.68 &    1588 &   34 \\
 12062 & J002556.06+153815.0 &  ...	     &0.0365 & 10.09 &  2.63 &  2.83 &  6.16 &  8.67 &  0.15 & 15.41 &   0.708 &   46 &  2.36 &     222 &   65 \\
\end{tabular}
\end{table*}

\setcounter{table}{1}
\begin{table*}
\small
\centering
\caption{-- {\it continued}}
\begin{tabular}{rclccccccccccccc}
\hline
   &  &  &  & Log \Mst & \Rinz & $R_{50}$ & $R_{90}$ & Log \must  & ext$_r$ & r & (b/a)$_r$ & incl &\nuvr & T$_{NUV}$ & \tmax \\
GASS  & SDSS ID & Other name & $z_{\rm SDSS}$ & (\Msun) & (\arcsec) & (\arcsec)&(\arcsec) & (\Msun~kpc$^{-2}$)& (mag)& (mag) &   & (deg) & (mag)& (sec) & (min) \\
(1)  & (2)  & (3)  & (4)  & (5) & (6) & (7) & (8)  & (9)  &  (10) & (11) & (12) & (13) & (14) & (15) & (16)\\
\hline
 51462 & J075600.62+141144.6 &  CGCG058-072  &0.0357 & 11.10 &  2.97 &  3.04 &  9.69 &  9.59 &  0.09 & 13.63 &   0.483 &   63 &  6.11 &     196 &    4 \\
 51336 & J075617.10+143609.6 &  ...	     &0.0474 & 11.00 &  5.38 &  6.18 & 13.64 &  8.72 &  0.09 & 14.66 &   0.732 &   44 &  4.20 &     209 &   19 \\
 19132 & J080020.05+222634.8 &  ...	     &0.0350 & 10.35 &  2.04 &  2.10 &  6.46 &  9.18 &  0.18 & 15.50 &   0.832 &   35 &  5.74 &    1696 &   54 \\
 56320 & J080342.27+100159.7 &  ...	     &0.0337 & 10.22 &  2.69 &  2.87 &  8.25 &  8.85 &  0.07 & 15.64 &   0.776 &   40 &  6.37 &     218 &   47 \\
 51580 & J080403.84+150518.4 &  ...	     &0.0390 & 10.24 &  2.24 &  2.49 &  7.46 &  8.90 &  0.10 & 16.52 &   0.264 &   80 &  4.27 &    2650 &   85 \\
\end{tabular}
\end{table*}

\setcounter{table}{1}
\begin{table*}
\small
\centering
\caption{-- {\it continued}}
\begin{tabular}{rclccccccccccccc}
\hline
   &  &  &  & Log \Mst & \Rinz & $R_{50}$ & $R_{90}$ & Log \must  & ext$_r$ & r & (b/a)$_r$ & incl &\nuvr & T$_{NUV}$ & \tmax \\
GASS  & SDSS ID & Other name & $z_{\rm SDSS}$ & (\Msun) & (\arcsec) & (\arcsec)&(\arcsec) & (\Msun~kpc$^{-2}$)& (mag)& (mag) &   & (deg) & (mag)& (sec) & (min) \\
(1)  & (2)  & (3)  & (4)  & (5) & (6) & (7) & (8)  & (9)  &  (10) & (11) & (12) & (13) & (14) & (15) & (16)\\
\hline
 26406 & J102149.72+132649.6 &  CGCG065-008  &0.0322 & 10.76 &  4.83 &  5.98 & 15.67 &  8.92 &  0.12 & 14.23 &   0.714 &   46 &  4.47 &     111 &   12 \\
 54240 & J102253.59+243623.0 &  IC2569       &0.0463 & 11.13 &  4.61 &  4.98 & 15.67 &  9.01 &  0.06 & 14.32 &   0.867 &   31 &  5.97 &     199 &    9 \\
 26503 & J102314.32+125224.0 &  ...	     &0.0329 & 10.45 &  1.92 &  2.00 &  6.46 &  9.39 &  0.11 & 15.15 &   0.528 &   60 &  5.36 &     322 &   42 \\
 26436 & J102413.51+131444.8 &  ...	     &0.0326 & 10.01 &  2.36 &  2.36 &  6.00 &  8.78 &  0.11 & 16.09 &   0.883 &   29 &  5.57 &     322 &   41 \\
 23029 & J102705.85+110317.5 &  IC612	     &0.0323 & 10.90 &  3.65 &  3.76 & 11.45 &  9.30 &  0.08 & 14.12 &   0.524 &   60 &  5.72 &     335 &    6 \\
\end{tabular}
\end{table*}

\setcounter{table}{1}
\begin{table*}
\small
\centering
\caption{-- {\it continued}}
\begin{tabular}{rclccccccccccccc}
\hline
   &  &  &  & Log \Mst & \Rinz & $R_{50}$ & $R_{90}$ & Log \must  & ext$_r$ & r & (b/a)$_r$ & incl &\nuvr & T$_{NUV}$ & \tmax \\
GASS  & SDSS ID & Other name & $z_{\rm SDSS}$ & (\Msun) & (\arcsec) & (\arcsec)&(\arcsec) & (\Msun~kpc$^{-2}$)& (mag)& (mag) &   & (deg) & (mag)& (sec) & (min) \\
(1)  & (2)  & (3)  & (4)  & (5) & (6) & (7) & (8)  & (9)  &  (10) & (11) & (12) & (13) & (14) & (15) & (16)\\
\hline
 48604 & J112746.27+265734.5 &  CGCG156-077  &0.0334 & 10.60 &  3.76 &  4.34 & 10.11 &  8.95 &  0.05 & 14.64 &   0.657 &   50 &  3.20 &      51 &   29 \\
  6015 & J113524.48+021627.3 &  ...	     &0.0289 & 10.23 &  2.33 &  2.92 &  7.75 &  9.13 &  0.08 & 16.40 &   0.371 &   71 &  4.46 &    1714 &   25 \\
 23761 & J113704.29+125535.7 &  IC2945       &0.0345 & 11.07 &  5.96 &  6.71 & 21.97 &  8.99 &  0.08 & 13.80 &   0.727 &   44 &  5.06 &     341 &    4 \\
 23739 & J113706.07+115237.7 &  ...	     &0.0358 & 10.90 &  2.72 &  2.93 &  9.64 &  9.47 &  0.08 & 14.58 &   0.432 &   67 &  6.13 &     252 &    9 \\
 23789 & J114144.66+122937.1 &  ...	     &0.0342 & 10.41 &  3.23 &  3.37 &  9.03 &  8.87 &  0.11 & 15.11 &   0.913 &   25 &  5.69 &     112 &   50 \\
\end{tabular}
\end{table*}

\setcounter{table}{1}
\begin{table*}
\small
\centering
\caption{-- {\it continued}}
\begin{tabular}{rclccccccccccccc}
\hline
   &  &  &  & Log \Mst & \Rinz & $R_{50}$ & $R_{90}$ & Log \must  & ext$_r$ & r & (b/a)$_r$ & incl &\nuvr & T$_{NUV}$ & \tmax \\
GASS  & SDSS ID & Other name & $z_{\rm SDSS}$ & (\Msun) & (\arcsec) & (\arcsec)&(\arcsec) & (\Msun~kpc$^{-2}$)& (mag)& (mag) &   & (deg) & (mag)& (sec) & (min) \\
(1)  & (2)  & (3)  & (4)  & (5) & (6) & (7) & (8)  & (9)  &  (10) & (11) & (12) & (13) & (14) & (15) & (16)\\
\hline
 44021 & J134231.07+301500.1 &  CGCG161-128  &0.0363 & 11.06 &  4.03 &  4.23 & 14.07 &  9.27 &  0.04 & 14.18 &   0.460 &   65 &  5.23 &     148 &    5 \\
 38018 & J134834.19+245329.2 &  ...	     &0.0297 & 10.08 &  4.18 &  5.09 & 13.46 &  8.44 &  0.04 & 16.24 &   0.230 &   83 &  3.35 &     178 &   28 \\
 35981 & J135308.35+354250.5 &  UGC8802      &0.0411 & 10.30 &  4.78 &  8.79 & 19.46 &  8.25 &  0.04 & 15.31 &   0.458 &   65 &  2.50 &    3864 &  106 \\
 44856 & J135411.14+243322.5 &  ...	     &0.0286 & 10.05 &  1.60 &  1.64 &  4.92 &  9.28 &  0.04 & 15.85 &   0.669 &   49 &  5.29 &     268 &   24 \\
 44892 & J135609.30+251143.6 &  CGCG132-055  &0.0290 & 10.66 &  4.42 &  4.78 & 15.64 &  9.00 &  0.05 & 14.40 &   0.769 &   41 &  5.59 &     268 &   12 \\
\end{tabular}
\end{table*}

\setcounter{table}{1}
\begin{table*}
\small
\centering
\caption{-- {\it continued}}
\begin{tabular}{rclccccccccccccc}
\hline
   &  &  &  & Log \Mst & \Rinz & $R_{50}$ & $R_{90}$ & Log \must  & ext$_r$ & r & (b/a)$_r$ & incl &\nuvr & T$_{NUV}$ & \tmax \\
GASS  & SDSS ID & Other name & $z_{\rm SDSS}$ & (\Msun) & (\arcsec) & (\arcsec)&(\arcsec) & (\Msun~kpc$^{-2}$)& (mag)& (mag) &   & (deg) & (mag)& (sec) & (min) \\
(1)  & (2)  & (3)  & (4)  & (5) & (6) & (7) & (8)  & (9)  &  (10) & (11) & (12) & (13) & (14) & (15) & (16)\\
\hline
  7813 & J151243.59+012752.2 &  CGCG021-049  &0.0293 & 10.79 &  4.15 &  4.74 & 15.08 &  9.17 &  0.13 & 14.04 &   0.581 &   56 &  5.60 &    1811 &    7 \\
 25057 & J152106.26+304036.9 &  ...	     &0.0308 & 10.01 &  3.12 &  2.90 &  8.04 &  8.59 &  0.06 & 15.81 &   0.719 &   45 &  5.35 &    2185 &   32 \\
 25115 & J152112.78+303928.5 &  CGCG165-039  &0.0308 & 10.50 &  2.93 &  2.79 &  8.96 &  9.14 &  0.06 & 14.60 &   0.537 &   59 &  5.40 &    2185 &   32 \\
 39407 & J152239.21+083226.7 &  ...	     &0.0366 & 10.26 &  2.57 &  2.64 &  7.44 &  8.86 &  0.10 & 15.89 &   0.484 &   63 &  5.26 &    1513 &   65 \\
 39532 & J152346.52+083853.1 &  ...	     &0.0301 & 10.24 &  4.03 &  4.09 & 11.06 &  8.62 &  0.09 & 15.55 &   0.731 &   44 &  6.09 &    1513 &   29 \\
\hline
\end{tabular}
\end{table*}
\end{landscape}


\onecolumn
\begin{table*}
\tiny
\caption{\hi\ Properties of GASS Detections.}
\label{t_det}
\begin{tabular}{cccccccccccccl}
\hline
      &         &               & $T_{\rm on}$ & $\Delta v$ &     & \whi  & \whi$^c$&  $F$      &  rms & &Log \Mhi  &   & \\
GASS  & SDSS ID & $z_{\rm SDSS}$ & (min)       &  (\kms)    & $z$ &  (\kms)& (\kms) & (Jy \kms) & (mJy)& S/N  & (\Msun) & Log \Mhi/\Mst & Q \\
(1)  & (2)  & (3)  & (4)  & (5)  & (6)  & (7)  & (8)  & (9)  &  (10) & (11) & (12) & (13) & (14)\\
\hline\hline
 11910 & J000632.57+154004.7 & 0.0370 &  14 &  13 & 0.037052 & 317$\pm$   1 & 300 &  0.45$\pm$  0.06 &  0.41 &  12.1 &  9.43 & $-$0.85 &  1    \\
 12062 & J002556.06+153815.0 & 0.0365 &   5 &  13 & 0.036445 & 255$\pm$  12 & 240 &  1.18$\pm$  0.09 &  0.64 &  22.9 &  9.84 & $-$0.25 &  1*   \\
  3305 & J005709.66+143906.6 & 0.0494 &   5 &  13 & 0.049501 & 403$\pm$   6 & 378 &  1.30$\pm$  0.13 &  0.75 &  16.9 & 10.15 & $-$0.56 &  1    \\
  3284 & J010253.84+141140.3 & 0.0397 &  53 &  16 & 0.039637 & 289$\pm$   7 & 271 &  0.23$\pm$  0.03 &  0.20 &  12.3 &  9.20 & $-$1.40 &  1    \\
  3666 & J011803.07+153224.4 & 0.0382 &  10 &  13 & 0.038173 & 295$\pm$   8 & 278 &  0.60$\pm$  0.07 &  0.49 &  14.1 &  9.59 & $-$0.66 &  1*   \\
\end{tabular}
\end{table*}

\setcounter{table}{2}
\begin{table*}
\tiny
\caption{-- {\it continued}}
\begin{tabular}{cccccccccccccl}
\hline
      &         &               & $T_{\rm on}$ & $\Delta v$ &     & \whi  & \whi$^c$&  $F$      &  rms & &Log \Mhi  &   & \\
GASS  & SDSS ID & $z_{\rm SDSS}$ & (min)       &  (\kms)    & $z$ &  (\kms)& (\kms) & (Jy \kms) & (mJy)& S/N  & (\Msun) & Log \Mhi/\Mst & Q \\
(1)  & (2)  & (3)  & (4)  & (5)  & (6)  & (7)  & (8)  & (9)  &  (10) & (11) & (12) & (13) & (14)\\
\hline
 43963 & J134142.40+300731.5 & 0.0370 &   5 &  13 & 0.037112 & 309$\pm$   9 & 292 &  1.14$\pm$  0.11 &  0.71 &  18.1 &  9.84 & $-$1.21 &  1    \\
 38018 & J134834.19+245329.2 & 0.0297 &   4 &  13 & 0.029694 & 344$\pm$   3 & 328 &  0.72$\pm$  0.11 &  0.73 &  10.7 &  9.45 & $-$0.63 &  1    \\
 35981 & J135308.35+354250.5 & 0.0411 &   5 &   5 & 0.041152 & 376$\pm$   3 & 359 &  3.45$\pm$  0.14 &  1.33 &  42.2 & 10.41 &    0.11 &  1    \\
 44856 & J135411.14+243322.5 & 0.0286 &  25 &  15 & 0.028646 & 301$\pm$  18 & 285 &  0.49$\pm$  0.05 &  0.29 &  17.8 &  9.25 & $-$0.80 &  1*   \\
 13618 & J135622.01+043710.6 & 0.0339 &  49 &  15 & 0.033857 & 226$\pm$  12 & 211 &  0.10$\pm$  0.03 &  0.21 &   6.0 &  8.72 & $-$1.48 &  3*   \\
\hline
\end{tabular}
\end{table*}


\onecolumn
\begin{table*}
\small
\centering
\caption{GASS Non-detections.}
\label{t_ndet}
\begin{tabular}{ccccccccc}
\hline
      &         &              &  $T_{\rm on}$ &  rms  & Log \Mhi$_{,lim}$ &  \\
GASS  & SDSS ID & $z_{\rm SDSS}$& (min)        &  (mJy)& (\Msun)  & Log \Mhi$_{,lim}$/\Mst & Note\\
(1)  & (2)  & (3)  & (4)  & (5)  & (6)  & (7) & (8)\\
\hline\hline
 11892 & J000200.82+150132.1 & 0.0357 &  46 &  0.21 &  8.76 &  $-$1.78  &  *   \\
 11903 & J000458.72+154018.2 & 0.0373 &  65 &  0.18 &  8.73 &  $-$1.44  &  *   \\
 12030 & J001842.68+151142.6 & 0.0372 &  34 &  0.26 &  8.88 &  $-$1.77  &  ... \\
  3157 & J003032.94+145635.4 & 0.0381 &  55 &  0.20 &  8.80 &  $-$1.77  &  *   \\
  3258 & J005316.95+160556.1 & 0.0393 &  68 &  0.17 &  8.74 &  $-$1.53  &  *   \\
\end{tabular}
\end{table*}

\setcounter{table}{3}
\begin{table*}
\small
\centering
\caption{-- {\it continued}}
\begin{tabular}{ccccccccc}
\hline
      &         &              &  $T_{\rm on}$ &  rms  & Log \Mhi$_{,lim}$ &  \\
GASS  & SDSS ID & $z_{\rm SDSS}$& (min)        &  (mJy)& (\Msun)  & Log \Mhi$_{,lim}$/\Mst & Note\\
(1)  & (2)  & (3)  & (4)  & (5)  & (6)  & (7) & (8)\\
\hline
 35497 & J125650.61+285547.4 & 0.0270 &  20 &  0.30 &  8.66 &  $-$1.78  &  *   \\
 35475 & J125935.67+283304.9 & 0.0253 &  15 &  0.38 &  8.71 &  $-$1.58  &  *   \\
 25213 & J131222.82+114339.5 & 0.0320 &  40 &  0.22 &  8.67 &  $-$1.45  &  *   \\
 26936 & J131525.21+152522.2 & 0.0266 &   5 &  0.59 &  8.94 &  $-$1.81  &  ... \\
 35659 & J134159.72+294653.5 & 0.0449 &   8 &  0.56 &  9.38 &  $-$1.76  &  ... \\
\hline
\end{tabular}
\end{table*}

\twocolumn


\end{document}